\documentclass[doi-false, twocolumn,aps,amsmath,nofootinbib,pre,floatfix,superscriptaddress]{revtex4-2}
\usepackage{graphicx}
\usepackage{hyperref}
\hypersetup{breaklinks=true}
\usepackage{amssymb}
\usepackage[utf8]{inputenc}
\usepackage[bottom]{footmisc}
\usepackage{afterpage}

\begin{document}
\title{Criticality and Griffiths phases in random games with quenched disorder}
\author{Marco A. Amaral}
\affiliation{Universidade Federal do Sul da Bahia - BA, Brazil}
\author{Marcelo M. de Oliveira}
\affiliation{Universidade Federal de São João Del Rey - MG, Brazil}

\date{\today}

\begin{abstract}
The perceived risk and reward for a given situation can vary depending on resource availability, accumulated wealth, and other extrinsic factors such as individual backgrounds. Based on this general aspect of everyday life, here we use evolutionary game theory to model a scenario with randomly perturbed payoffs in a prisoner's dilemma game. 
The perception diversity is modeled by adding a zero-average random noise in the payoff entries and a Monte-Carlo simulation is used to obtain the population dynamics. 
This payoff heterogeneity can promote and maintain cooperation in a competitive scenario where only defectors would survive otherwise.
In this work, we give a step further understanding the role of heterogeneity by investigating the effects of quenched disorder in the critical properties of random games. We observe that payoff fluctuations induce a very slow dynamic, making the cooperation decay behave as power laws with varying exponents, instead of the usual exponential decay after the critical point, showing the emergence of a Griffiths phase. We also find a symmetric Griffiths phase near the defector's extinction point when fluctuations are present, indicating that Griffiths phases may be frequent in evolutionary game dynamics and play a role in the coexistence of different strategies.

\end{abstract}


\maketitle

\section{Introduction}
\label{intro}

A wealthy person can perceive the risk of losing a car in a bet as a minor nuisance, while the same situation could be viewed as a huge loss for a less fortunate individual. The reward and risk perception of the same situation can greatly vary from one person to another, depending on many factors such as accumulated wealth, food availability, psychological situation, and so on \cite{wang_z_pre14b, Stewart2016}.
Evolutionary Game Theory (EGT) \cite{Nowak2006, Szabo2007} has been one of the most successful frameworks to model rational decision-making in conflicting situations. Its many applications range from economics \cite{jiang_ll_pone13} to epidemiology \cite{Amaral2020d, Jentsch2021, Kabir2021}, rumor spreading \cite{Amaral2018b}, quantum mechanics \cite{Vijayakrishnan2020} and even the evolution of moral behavior \cite{Kumar2020}.  Yet, a common assumption in this framework is that all individuals during a game share the same perceptions of the reward and risk, in terms of absolute values. This is a reasonable hypothesis when trying to simplify all the complexities of human and animal interactions, but important and subtle effects may be left out when using such assumption \cite{Amaral2020a, Amaral2020, tanimoto_pre07b, Perc2006, Perc2006b}.

In the context of EGT, one of the most long-standing questions is how cooperation can emerge in a competitive scenario \cite{Pennisi2005, galam_ijmpc08, Capraro2018, rand_tcs13}. A lot of effort have been dedicated into uncovering which mechanisms may promote cooperation  ~\cite{rand_tcs13, Perc2017,buchan_pnas09, Gomez-Gardenes2007, Szabo2016a, galam_ijmpc08, Capraro2018, roca_plr09, Nowak2006, Szabo2007, perc_bs10}. 
Among the most famous, we have kin selection~\cite{Hamilton1964}, direct and indirect reciprocity \cite{trivers_qrb71, axelrod_s81}, network reciprocity~\cite{Nowak1992a, wardil_epl09, wardil_pre10, wardil_jpa11, NagChowdhury2020, Vukov2012, Wu2018a} and group selection~\cite{wilson_ds_an77}. Specifically, heterogeneity (sometimes deemed as diversity) have recently gained a lot of interest as another mechanism that allows emergent phenomena to help increase cooperation \cite{Zhao2020, Sendina-Nadal2020, santos_jtb12, Sparrow1999, fort2008minimalevol, Szolnoki2018a, perc_bs10, Amaral2016, Amaral2015, Tanimoto2017}.

While traditional EGT has provided fundamental models and methods that enable us to study the evolution of cooperation, the complexity of such systems also requires methods of non-equilibrium statistical physics to be used to better understand the emergence and dynamics of cooperation, and also to reveal the hidden mechanisms that promote it \cite{Perc2016a}.
The effects of disorder in non-equilibrium phase transitions have been an important topic of research in statistical physics in the last decades \cite{Marrobook, odorbook, henkel08}. Both quenched (frozen) \cite{adr-dic98, vojta06, DeOliveira2008, vojta09, Gonzaga2019} as well as time-dependent (temporal/annealed) disorder \cite{TGP, Barghathi2016, temporal1, temporal2, Fiore2018} have provided rich phenomena and phase diagrams. 
Depending on the universality class of the non-disordered (clean) model, disorder can be a relevant perturbation, changing the critical exponents, and exotic phases with unusual scaling can emerge \cite{harris74, janke, vojta14, Girardi-Schappo2016}. E.g., in models belonging to the directed percolation (DP) universality class, such as the imitation dynamics for an evolutionary game \cite{Hauert2005, Szabo2007}, quenched uncorrelated randomness may produce rare regions which are locally supercritical even when the whole system is sub-critical \cite{vojta14}. 
Those have been observed in magnetic systems \cite{Bray1987, magnetic} and epidemic dynamics \cite{DeOliveira2008, epidemics, Odor2015, Cota2016} for example, but not in evolutionary game systems until now.
The lifetime of such ``active rare regions'' grows exponentially with the domain size, usually leading to slow dynamics, characterized by non-universal exponents towards the extinction, for some interval of the control parameter. This interval of singularities is called Griffiths phase ~\cite{rojas-echenique_e11, Bray1987, Girardi-Schappo2016}.

As in nature, clean systems are more of an exception than the rule, and in real social systems, heterogeneity is an unavoidable ingredient. On the other hand, most of the studies in EGT do not focus on the temporal dynamics towards the stationary dominant state. In the present work, we aim to provide a detailed investigation of such dynamics as well as to understand the role of the heterogeneity in the population.

In particular, here we model the diversity of perceptions between individuals by introducing perturbations in the payoff matrix of a two-player Prisoner's Dilemma game. The use of payoff perturbations (some times deemed as multi-games, random games, or stochastic games) has recently attracted a lot of attention since it describes a common phenomenon regarding perception diversity ~\cite{Szolnoki2019, Takesue2019, Wu2018, Zhou2018, Su2019, perc_pre08, Qin2017a, Stollmeier2018, Alam2018, Hilbe2018, Tanimoto2016, Yakushkina2015, wang_z_pre14b, Zhang2013, anh_tpb12, Perc2006, Perc2006a}.
Following the work initially done in \cite{Amaral2020a, Amaral2020}, we use small, zero-average, random perturbations in the agent's payoff to better understand how said fluctuations can affect the dynamics of a population modeled by evolutionary game theory.
We will focus on analyzing the critical properties and the emergent phase that occur near the phase transitions when there is disorder in the payoff structure. While heterogeneity has been shown to be a strong promoter of cooperation in competitive games, the exotic phases that may appear in such states are up to now not well understood. We study such states in the light of statistical physics, looking for general properties that indicate the emergence of a Griffiths phase when there is perturbation on the system.

\section{The model} 
\label{model}

We consider pairwise, two-strategy games whose agents can either cooperate (C) or defect (D). Mutual cooperation yields a payoff $R$ (reward), while mutual defection yields $P$ (punishment). If one player cooperates with a defector, the defector receives a payoff $T$ (temptation) while the cooperator receives a payoff $S$, known as the Sucker's payoff~\cite{Szabo2007}.
We model the perception diversity as small random perturbations in the payoff values, as done in \cite{Amaral2020a,Amaral2020}. Each payoff entry is independently perturbed with a random value with zero average, as we want the perturbations to be symmetrical and not favor any specific strategy on average.
%
%
We denote the perturbations as $\varepsilon$, where they are drawn from a uniform distribution with range $\Delta$ (e.g. $-\Delta < \varepsilon < \Delta$) and they are not cumulative. Here, $\Delta$ is the control parameter that gives the perturbation strength.
In general, the payoff matrix ($G$) is denoted as:
\begin{equation}\label{paymatrix}
\begin{array}{c c} &
\begin{array}{c c} C~~ & ~~D \\
\end{array}
\\
\begin{array}{c c}
C \\
D
\end{array}
&
\left[
\begin{array}{c c }
R+\varepsilon _R & S+\varepsilon _S \\
T+\varepsilon _T & P+\varepsilon _P
\end{array}
\right]
\end{array}
\end{equation}

\begin{figure}
  \includegraphics[width=5.7cm]{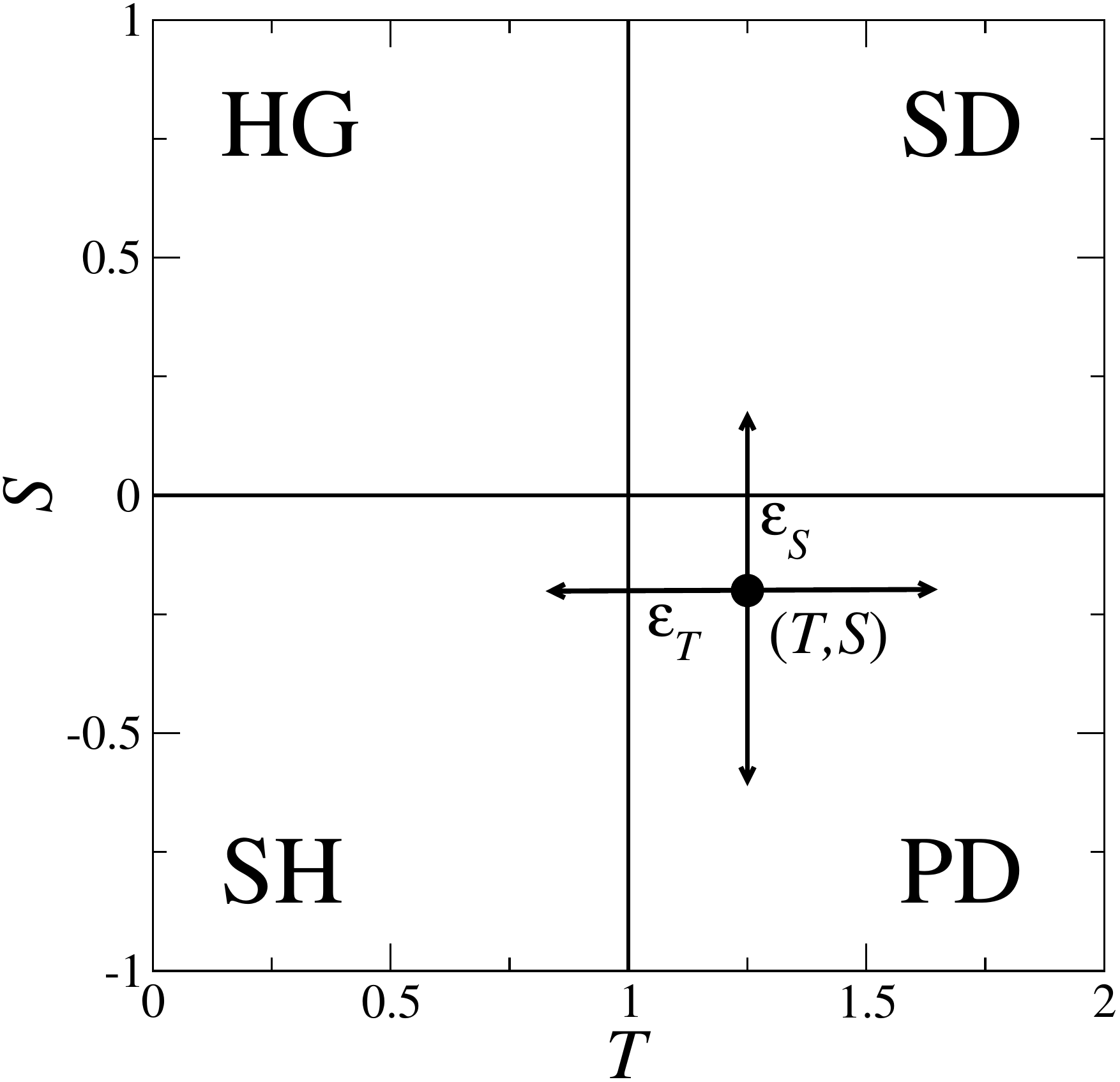}
  \caption{$T\times S$  parameter space with $R=1,P=0$, spanning four game classes. Fluctuations can act over $[T,S,P,R]$ simultaneously and uncorrelated. The payoff matrix has parameters that fluctuate around a 4-dimensional box with $2\Delta$ size edges, centered on the original game. Note that the perturbations may lead agents to (locally) play different classes of games.}
  \label{diagram}
\end{figure}

We set $R=1$ and $P=0$ without loss of generality \cite{Szabo2007, perc_bs10}. This allow us to organize the main classes of dilemma games in a $T \times S$ parameter diagram, with $T\in[0,2]$ and $S\in[-1,1]$, as can be seen in Figure \ref{diagram}. We delimit four quadrants with the Harmony Game (HG), Stag-Hunt (SH), Snow-drift (SD), and Prisoner's Dilemma (PD).
%

Regarding the perturbations, previous works \cite{Amaral2020a, Amaral2020, tanimoto_pre07b, Perc2006, Perc2006b} have shown that perturbations on the payoff matrix main diagonal may lead to slightly different results when compared with the matrix off-diagonal, depending on how the disorder is applied. 
We chose to perturb all four payoff entries, since initial simulations showed that, for our model, the main aspects of the Griffiths phase in a quenched disorder setting are more evident when all entries are perturbed. 
Additionally, here we focus on the quenched disorder, i.e. a `frozen' disorder fixed in time \cite{adr-dic98,vojta06,DeOliveira2008,vojta09,Gonzaga2019}. The perturbation on the payoff matrix is done only once, at the beginning of the simulation for every single agent, and remains fixed for the rest of that simulation. That means that each site $i$ will have its own perturbed matrix $G_i$, which will not change over time. Note that the payoff matrices of two sites will have different values of perturbations, as expected since we model different perceptions of risk and reward.
On the other hand, the annealed disorder corresponds to a temporal disorder and initial results did not found traces of Griffiths phases utilizing this approach, therefore we did not explore this setting deeply. Nevertheless, we stress that other systems with some kinds of temporal disorder can show a more exotic ``temporal Griffiths Phase'' \cite{Fiore2018, Fiore2018, Vazquez2011}, and this may be also the case for temporal perturbations in game theory. Since the scope of the current work is focused on the usual Griffiths phase, we let this analysis for future works.

For the population dynamics, we implement the usual Monte-Carlo protocol with an imitative update rule weighted by the Fermi distribution~\cite{Szabo2007, perc_bs10, Javarone2018}, in a spatially distributed population with the square lattice topology and periodic boundary conditions. For the population update, first a player $i$ accumulates its payoff by playing against its four nearest neighbors (Von Neumann neighborhood). Next, $i$ updates its strategy by comparing its payoff with one randomly chosen neighbor, $j$ (we also obtain $j$'s payoff by making it play against all its nearest neighbors). Agent $i$ adopts the strategy of agent $j$ with probability
\begin{equation}\label{imitateeq}
P(u_i,u_j)=\frac{1}{ 1+e^{-(u_{j}-u_{i})/k} },
\end{equation}
\noindent where $k$ is the irrationality level \cite{Szabo2007}, and $u_{i}$ represents the payoff of agent $i$. We set $k=0.1$ for all simulations. One Monte-Carlo Step (MCS) is comprised of $N$ repetitions of this unitary update procedure, where $N$ is the number of agents in the population.
We run the simulations for at least $10^4$ MCS's for the system to reach equilibrium, but this number can increase considerably near the phase transition or during a Griffiths phase. After the equilibrium, we average the fraction of cooperators over $1000$ steps.  We used lattices of linear size $L=200$ and repeat this procedure for $100-200$ different simulation runs (samples) to obtain more accurate averages. Regarding the system size, we have also considered linear sizes varying from $L=100$ up to $L=1000$ agents. We observed that the behaviour is almost identical for sizes $L>100$, with minor quantitative deviations only for $L = 100$.

\section{Results}\label{results}

We begin by presenting the general effects of the perturbation on the population. 
We remind that a study of the general benefits of random payoffs to cooperation can be found in the references \cite{Amaral2020a, Amaral2020, tanimoto_pre07b, Perc2006, Perc2006b}. Here we shall focus more on the phase transition points and the Griffiths phase.

\begin{figure}
  \includegraphics[width=8cm]{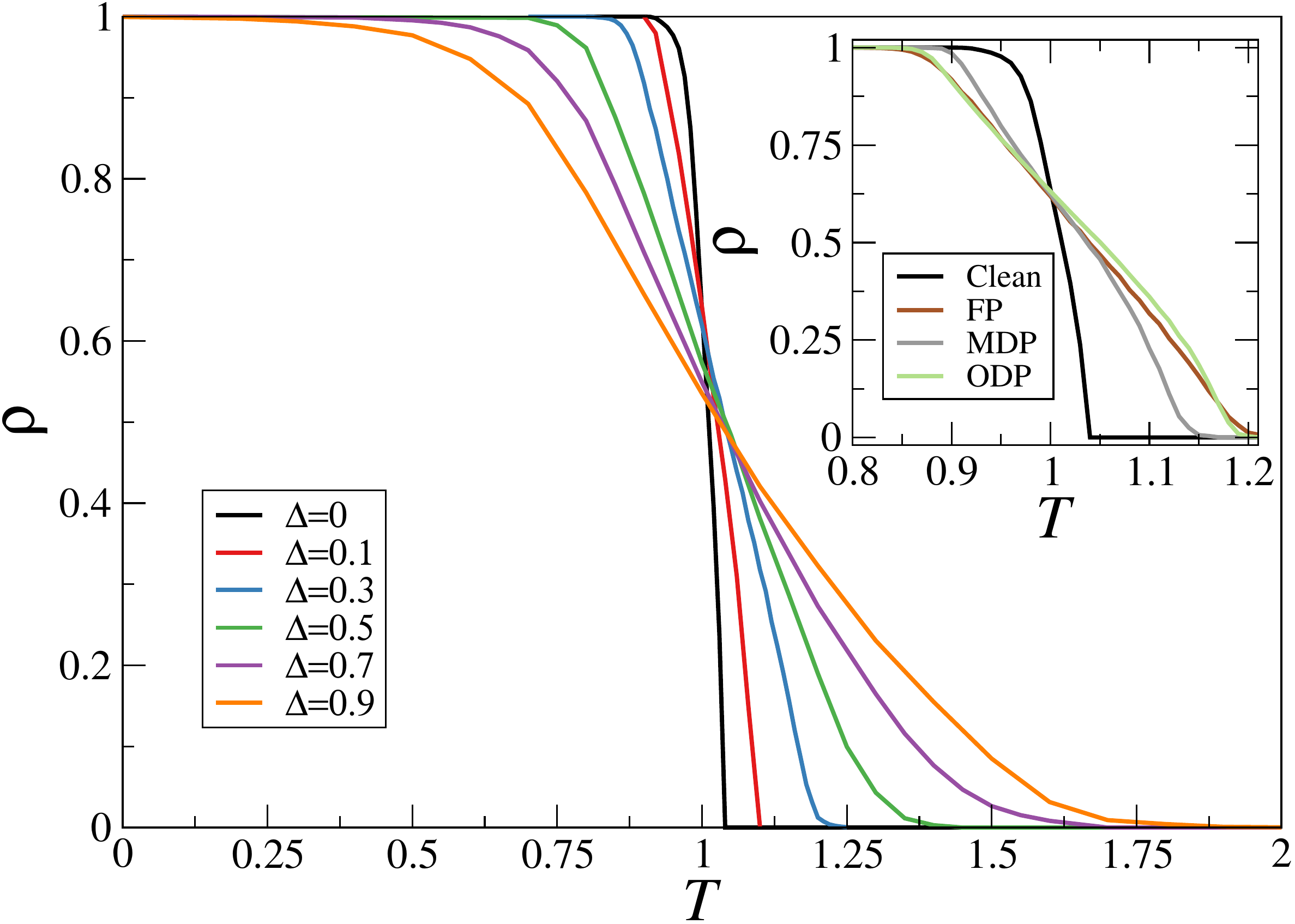}
  \caption{General effects of the perturbation in the cooperation levels. The graph shows the final cooperator fraction, $\rho$, as a function of the temptation to defect, $T$, for various perturbation strengths $\Delta$ in the weak Prisoner's Dilemma ($S=0$). Payoff perturbation can continuously increase cooperation for $T>1$, and the effect is reversed for $T<1$.
  For the clean model ($\Delta=0$) cooperation is extinct around $T_c=1.036$. The payoff perturbation increases the survival range of cooperation in the $T$ parameter as $\Delta$ increases.
  %
  {\em Inset:} Similar figure but comparing the clean and quenched perturbation model for different perturbation settings. We set $\Delta=0.3$ and show the Full Perturbation (FP), Main-Diagonal Perturbation (MDP) and Off-Diagonal Perturbation (ODP). We will focus mainly on the Full Perturbation model.%
}
\label{phasetransiall}
\end{figure}

In Fig. \ref{phasetransiall} we show the usual behavior of the cooperator's fraction, $\rho$, versus the temptation to defect, $T$, for various perturbation strengths $\Delta$ in the weak Prisoner's Dilemma ($S=0$). We chose the weak Prisoner's Dilemma scenario so as to better focus our attention of the Griffiths phase in a classical game setting. Note however that the obtained results strongly indicate that such phases may emerge in other games near the phase transition points. We stress that the perturbation is supposed to represent a small deviation in the risk and reward perception of each player. In this sense, we expect that reasonable values of the perturbation strength would be around $0<\Delta<0.5$, as the maximum payoff is $T=2$ in this parametrization. The main effect of the payoff perturbation is to continuously increase the cooperation for $T>1$ when compared to the clean model (zero perturbation). Also, note that in the region of the Harmony Game ($T<1$) this effect inverses, and cooperation diminishes with the perturbation.
The inset in Fig. \ref{phasetransiall} presents $\rho$ as a function of $T$ for a fixed perturbation strength $\Delta=0.3$, and shows the results for three possible payoff perturbations, i.e.  Full Perturbation (FP), Main-Diagonal Perturbation (MDP), and Off-Diagonal Perturbation (ODP). While the settings can present small differences, here we will focus on the FP model, since it is the one with stronger disorder effects, that are usually associated with the emergence of a Griffiths Phase.

\begin{figure*}
\includegraphics[width=8cm]{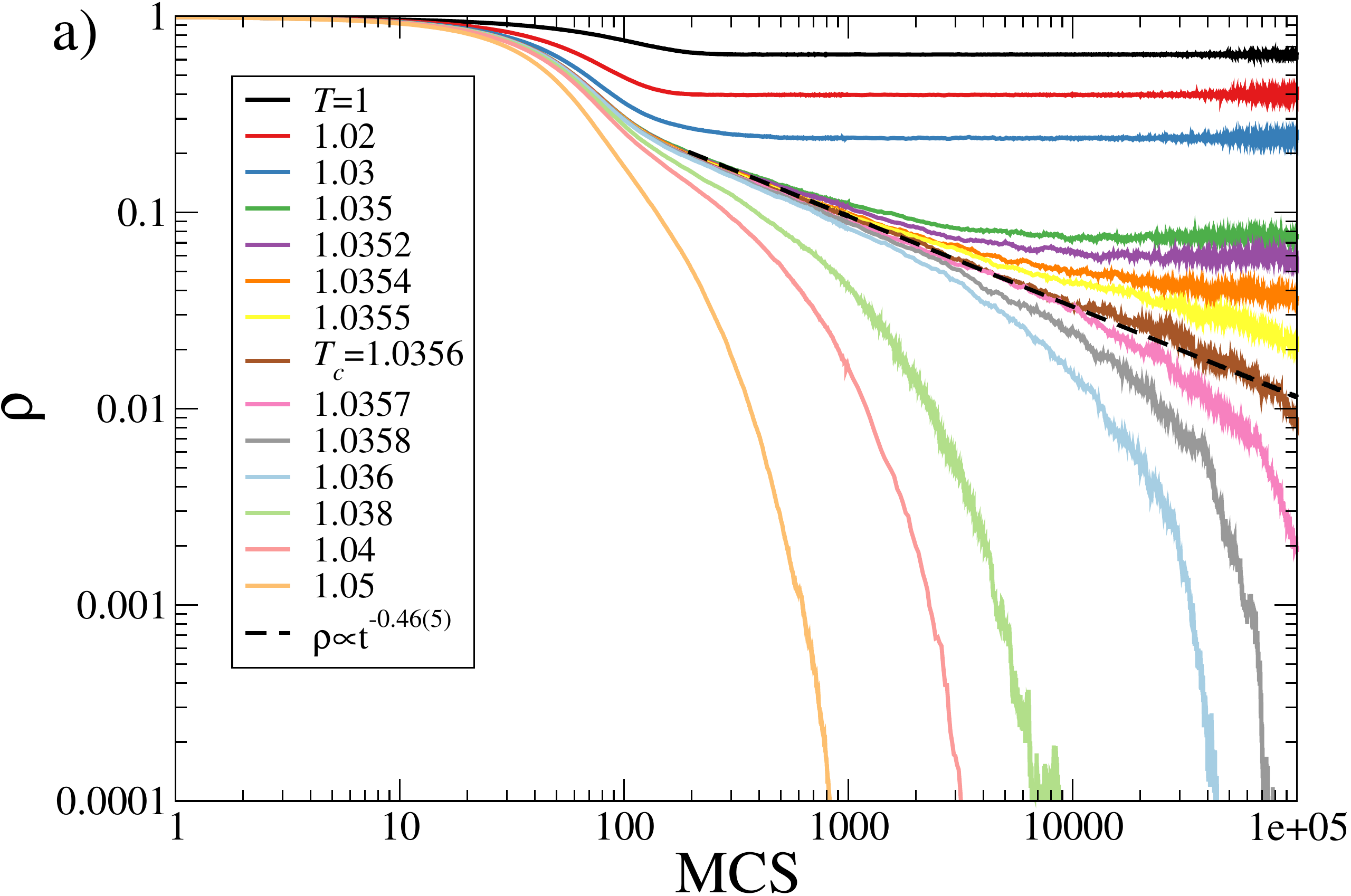}
\includegraphics[width=8cm]{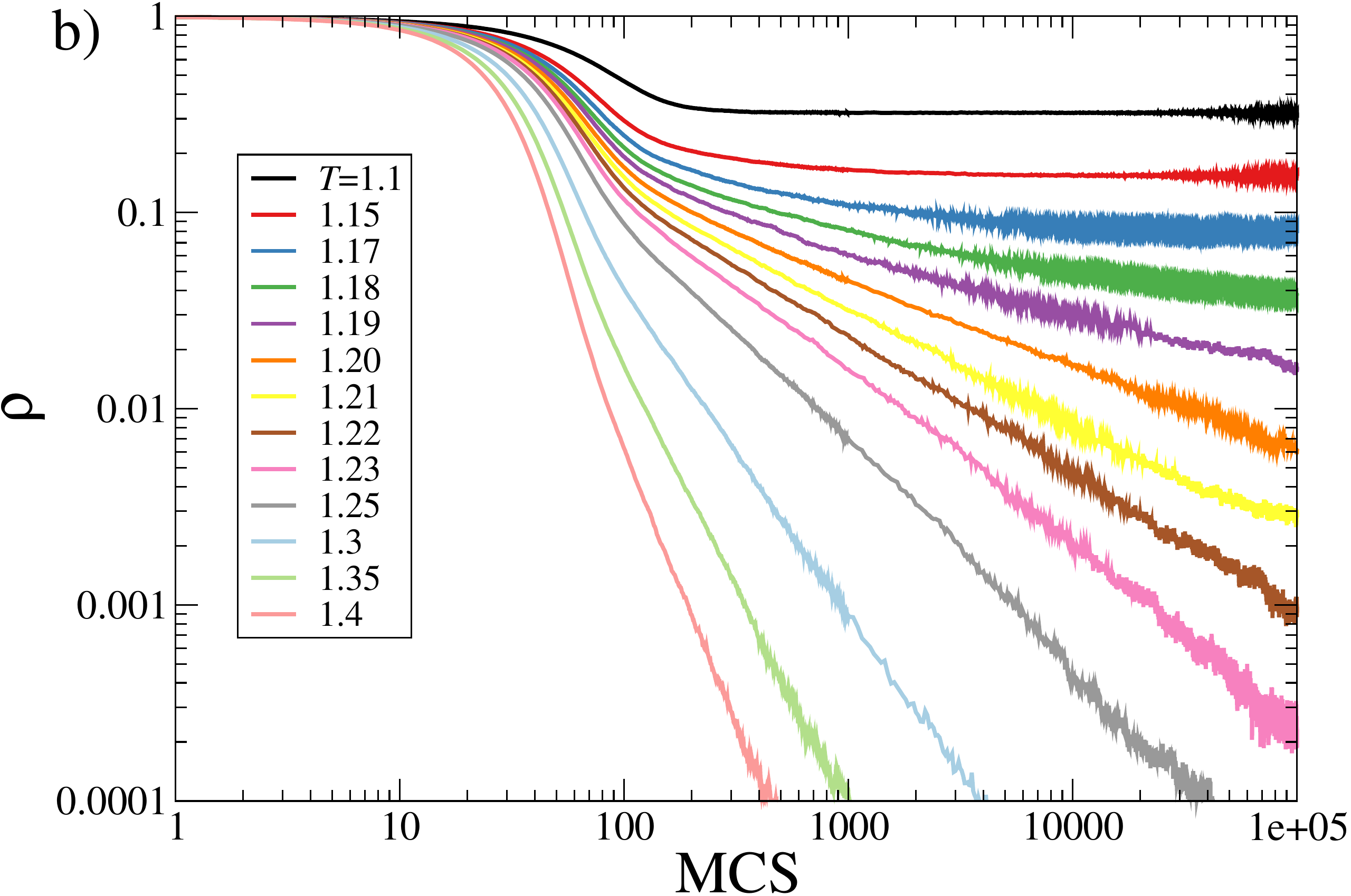}
  \caption{Decay dynamics for the cooperation fraction ($\rho$) versus Monte Carlo Steps (MCS), of the population using different $T$ values near the cooperation extinction point for the clean ($\Delta=0$), a), and perturbed ($\Delta=0.3$), b), dynamics. The simulations start with $99\%$ of the lattice occupied by cooperators, so as to better observe when the decay becomes a power-law for the perturbed model. For the clean model we can see that the system transitions from a stable state to an exponential decay. The point that marks the transition of the two regimes behaves as a power law, with $\rho\propto t^{-\delta} ; \delta=0.46(5)$. The perturbed model presents power law decay for a range of $T$ values, with varying decay exponents.}
\label{decaydynamics}
\end{figure*}

Now, we investigate the temporal dynamics via decay simulations (also called seed simulations) \cite{henkel08}, a common tool from nonequilibrium statistical physics to obtain important features during a phase transition. To do so, we run simulations starting with $99\%$ of the lattice filled with cooperators, in a region of high $T$ where cooperation should not survive
\footnote{We stress that usually, in the context of EGT, population dynamics is done with homogeneous starting conditions, i.e. both strategies start with equal densities. Nevertheless, such an approach is less useful for characterizing the power-law decay of a given strategy during a Griffiths phase. We note however that the simulations with homogeneous conditions were run and presented similar results regarding the final fraction of cooperators.}. 
Our results are shown in Fig. \ref{decaydynamics}, presenting  the evolution of cooperators for the clean (\ref{decaydynamics}a) and perturbed (\ref{decaydynamics}b) models near their respective phase transitions where cooperation is extinct. Here we use $\Delta=0.3$ for the perturbed model, but the general behavior is consistent for $0<\Delta<1$.

Looking at Figure ~\ref{decaydynamics}a), the clean model exhibits three distinct behaviors. Before the critical point $T<T_c\simeq1.0356(4)$, cooperators stay stable in the long term of the temporal evolution (at these values of the control parameter, $T$, the system is in the supercritical regime). For $T>T_c$ there is an exponential decay, with cooperation quickly reaching extinction (subcritical regime). Finally, at criticality (exactly at $T=T_c$) one observes a power-law decay $\rho\propto t^{-\delta}$. We find the exponent $\delta= 0.46(5)$, in agreement with the value $\delta=0.4505(10)$ exhibited by $2+1$ dimension models in the directed percolation (DP) universality class \cite{Marrobook}.

On the other hand, the decay dynamics of the disordered model present a different behavior. It is characterized by an activated dynamic scaling near the critical value of $T$ where cooperation is extinct,  showing a range of $T$ values where we observe a power-law decay (with diverse inclinations depending on the $T$ value and perturbation strength $\Delta$). This is a typical behavior of a Griffiths phase. Note that in Figure (\ref{decaydynamics}a)  we present a range of $1.0<T<1.05$  for the clean model, where $T_c\simeq1.0356(4)$ whereas in Figure (\ref{decaydynamics}b) we show the perturbed version using $1.1<T<1.4$.
Let $T_g$ be the temptation value for the perturbed model where cooperation begins to decay as a power law, indicating a Griffiths phase. In this regime, cooperation should tend to zero for infinite times, but at a much slower rate than the exponential decay of the clean model after the critical point $T_c$. We see that for $\Delta=0.3$, this value is around $T>T_g\simeq 1.18(1)$.

An interesting point to note is that the clean and perturbed dynamics have different asymptotic values of cooperation even before the Griffiths phase, that is, $T_c<T<T_g$. In this region of $T$, the Griffiths phase does not appear, the time evolution of the perturbed model behaves in a supercritical regime, with $\rho$ being stable. Even so, cooperation is still increased by the perturbation when compared with the clean model.
%
%
%
The effect of disorder is to create a Griffiths phase for high values of $T$. Nevertheless, for  $T_c<T<T_g$, the disorder is still able to promote cooperation in a region where it should have been extinct.

\begin{figure*}	
  \includegraphics[width=2.8cm]{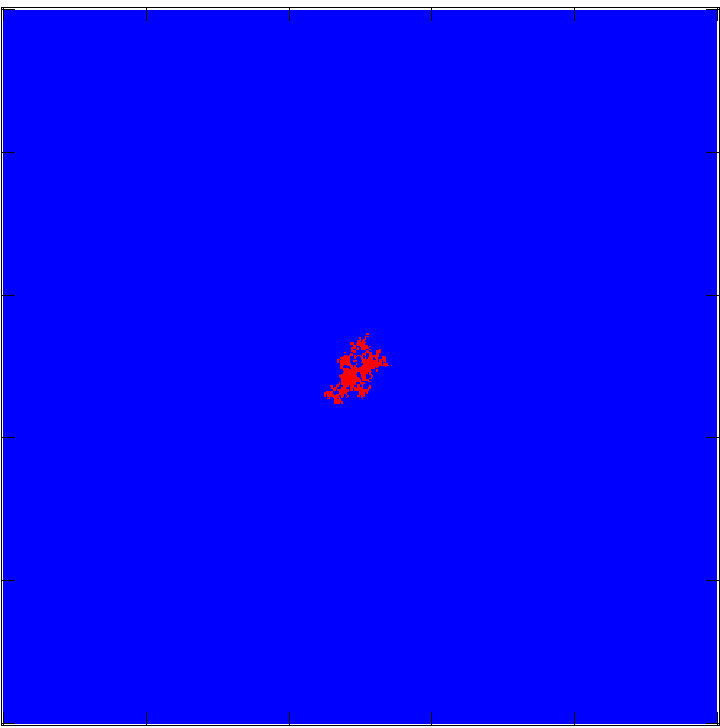}	
  \includegraphics[width=2.8cm]{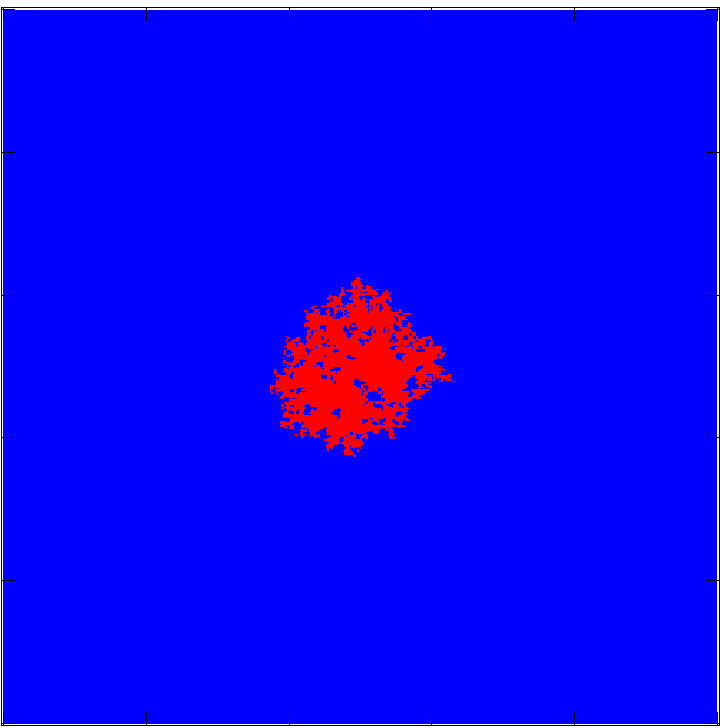}	
  \includegraphics[width=2.8cm]{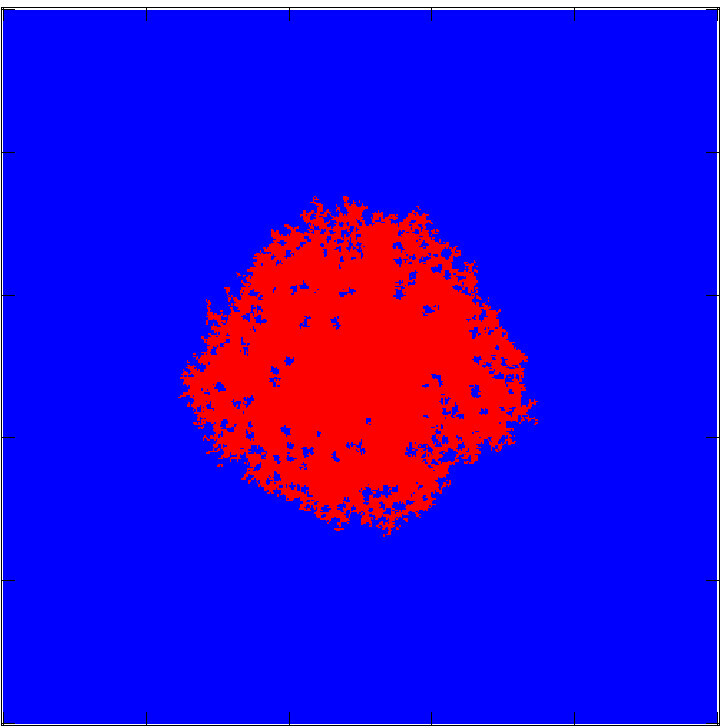}
  \includegraphics[width=2.8cm]{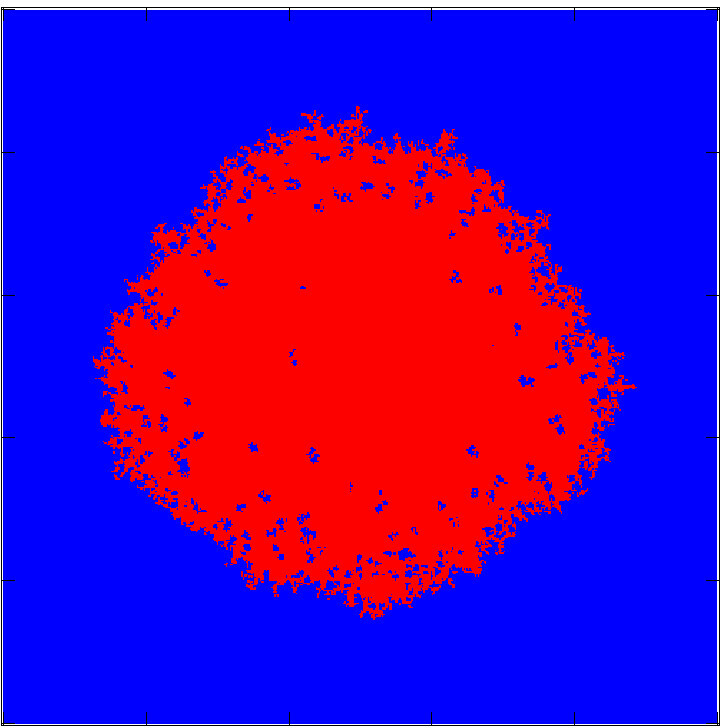}
  \includegraphics[width=2.8cm]{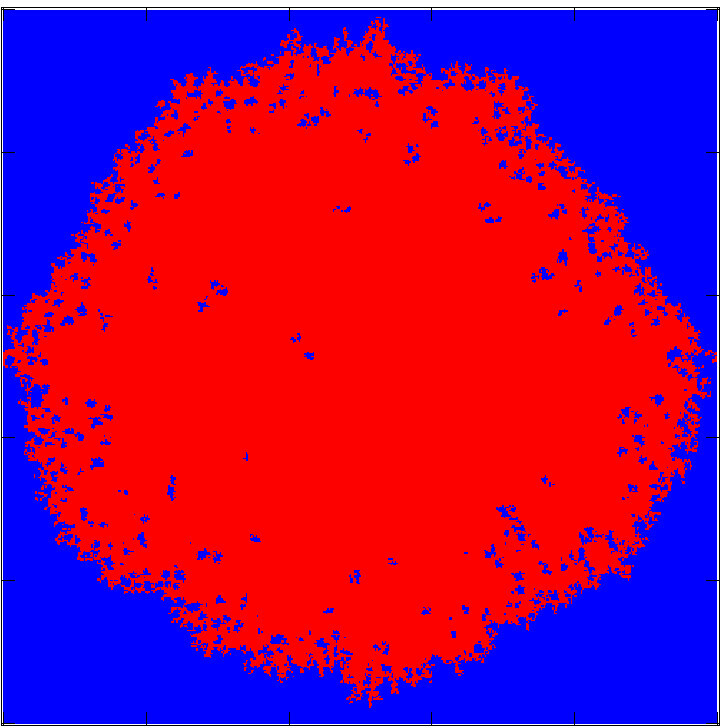}
  \includegraphics[width=2.8cm]{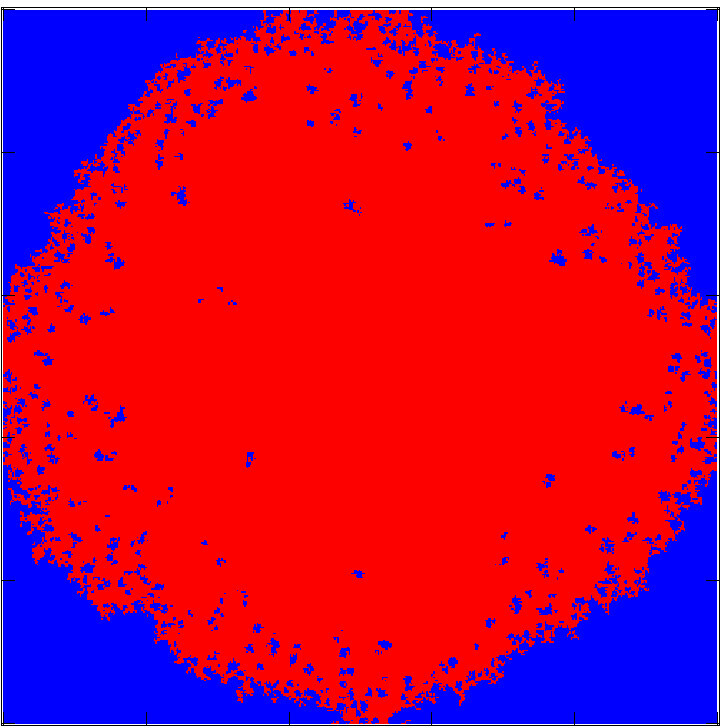}
  
  \includegraphics[width=2.8cm]{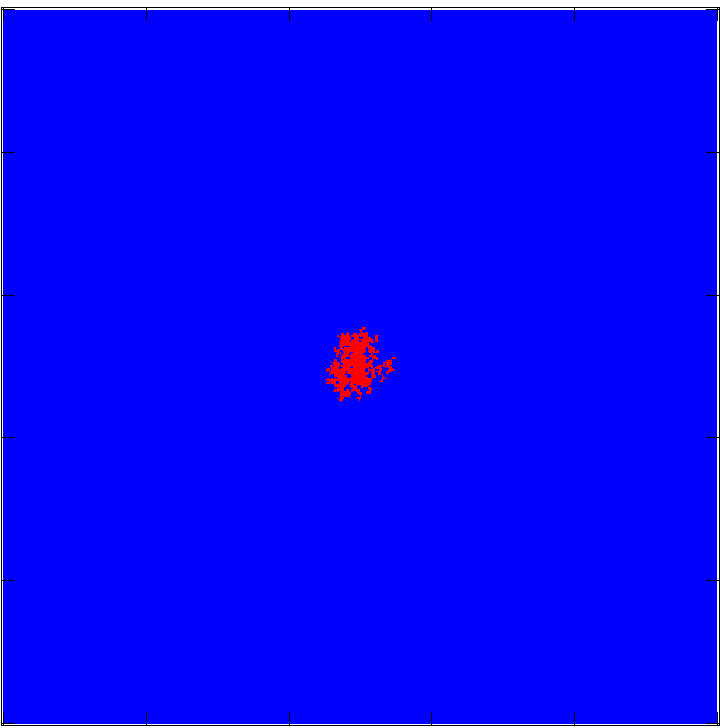}	
  \includegraphics[width=2.8cm]{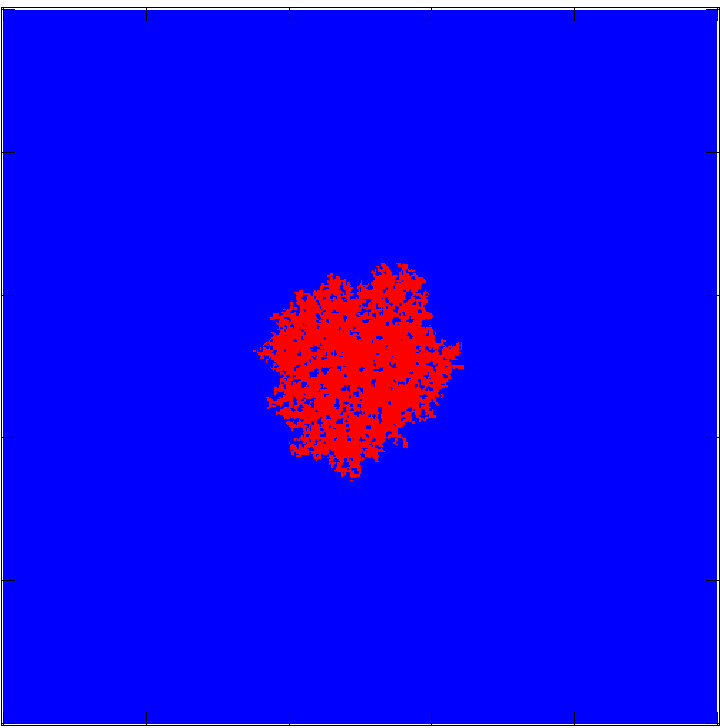}	
  \includegraphics[width=2.8cm]{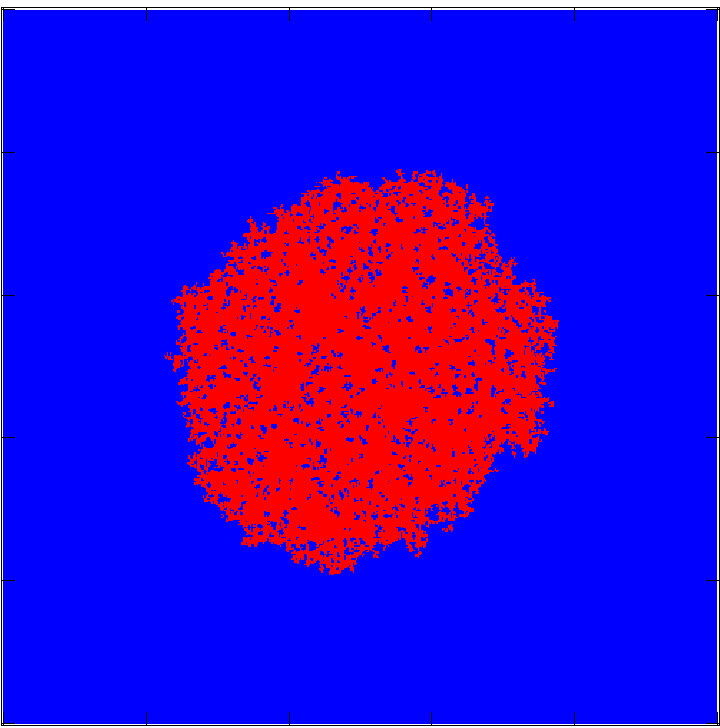}
  \includegraphics[width=2.8cm]{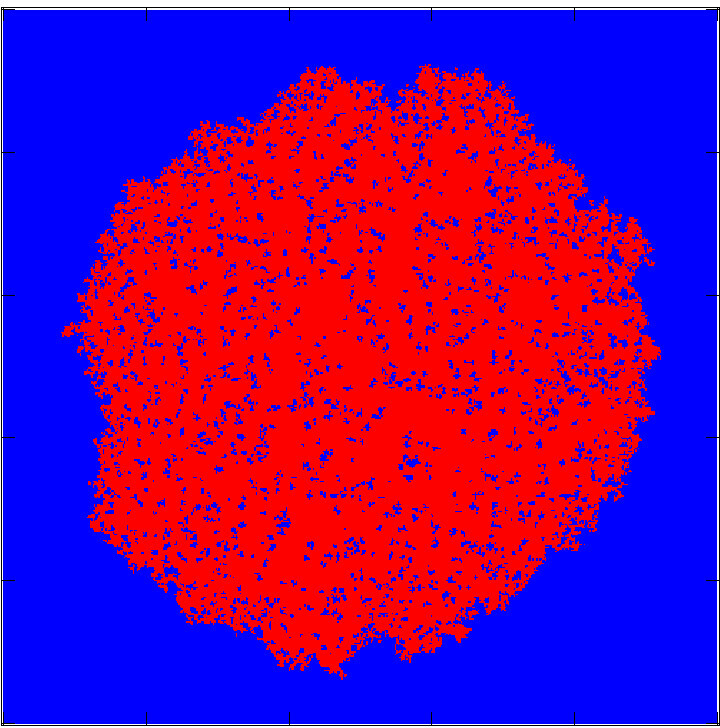}
  \includegraphics[width=2.8cm]{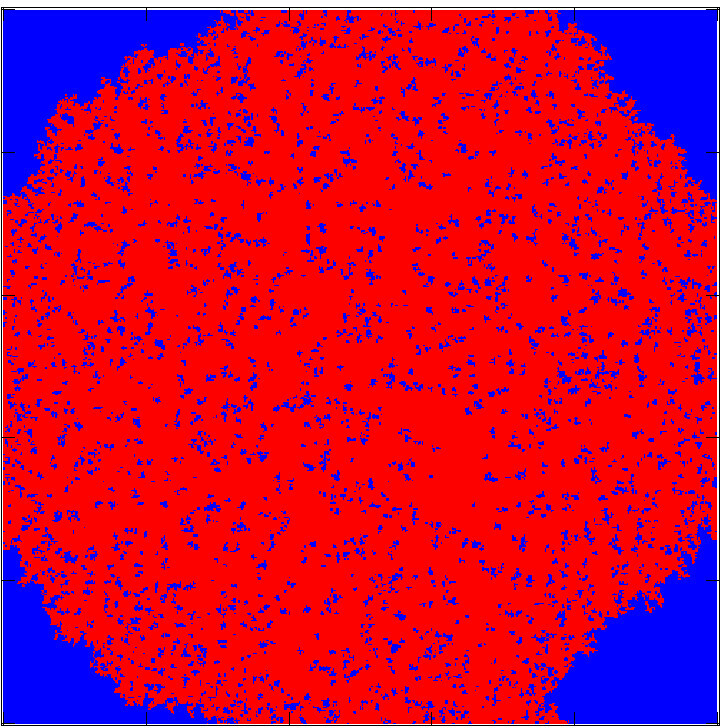}
  \includegraphics[width=2.8cm]{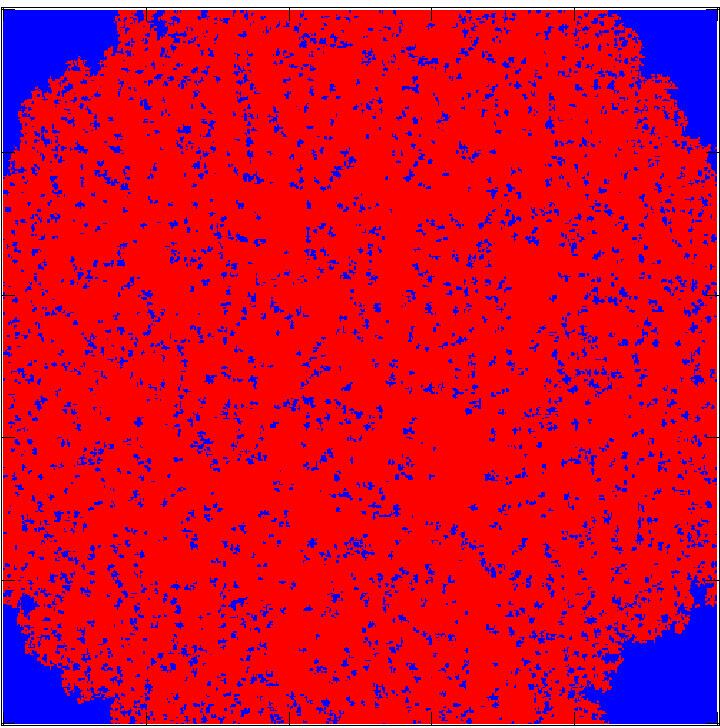}
    \caption{Lattice snapshots for the evolution of the clean (top) and perturbed (bottom) models. We use prepared initial conditions with just a few defectors (light red) in the center surrounded by cooperators (dark blue). The $T$ values are chosen such that cooperation will tend to zero in both cases. We can see that even if cooperation should be extinct in the long run, the perturbed model allows the survival of cooperative clusters in a meta-stable state for long periods, leading to the slow power law decay of cooperation. Here $T=1.046$ for the clean model and $T=1.364$ for the perturbed model with $\Delta=0.6$.}
  \label{snaps}	
\end{figure*}

We can also visualize the decay dynamics looking at the lattice snapshots. Figure \ref{snaps} presents the sequential snapshots of a clean (top row) and perturbed model (botton row) in a specially prepared initial condition. The lattice begins the simulation with only a small cluster of defectors surrounded by cooperators. Here we use $\Delta=0.6, T=1.364$ for the perturbed model and $T=1.046$ for the clean model. The $T$ values are chosen so that cooperation should be extinct in both models for long times. The main aspect to note here is how cooperator clusters are able to survive for much longer times in the perturbed model. This can be seen by looking at the expansion border of the defectors. In both models we see small cooperator clusters near the border, but only in the perturbed model that those clusters are able to survive for longer times (due to the Griffths phase). Note that eventually all clusters will succumb to defection, but this will be a very slow decay as a power law, and not an exponential decay.

\begin{figure*}	
 \includegraphics[width=8cm]{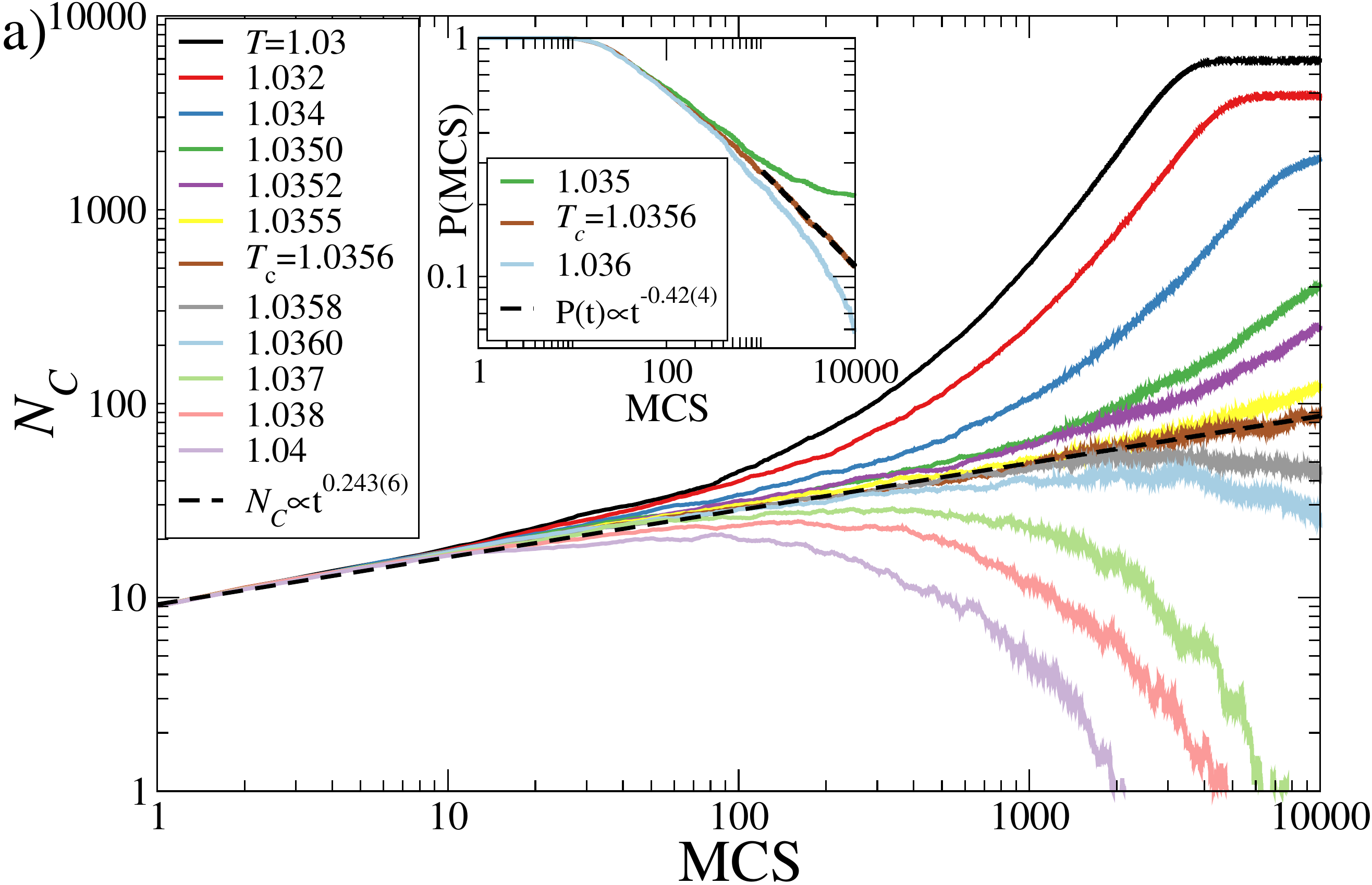} 	
    \includegraphics[width=8cm]{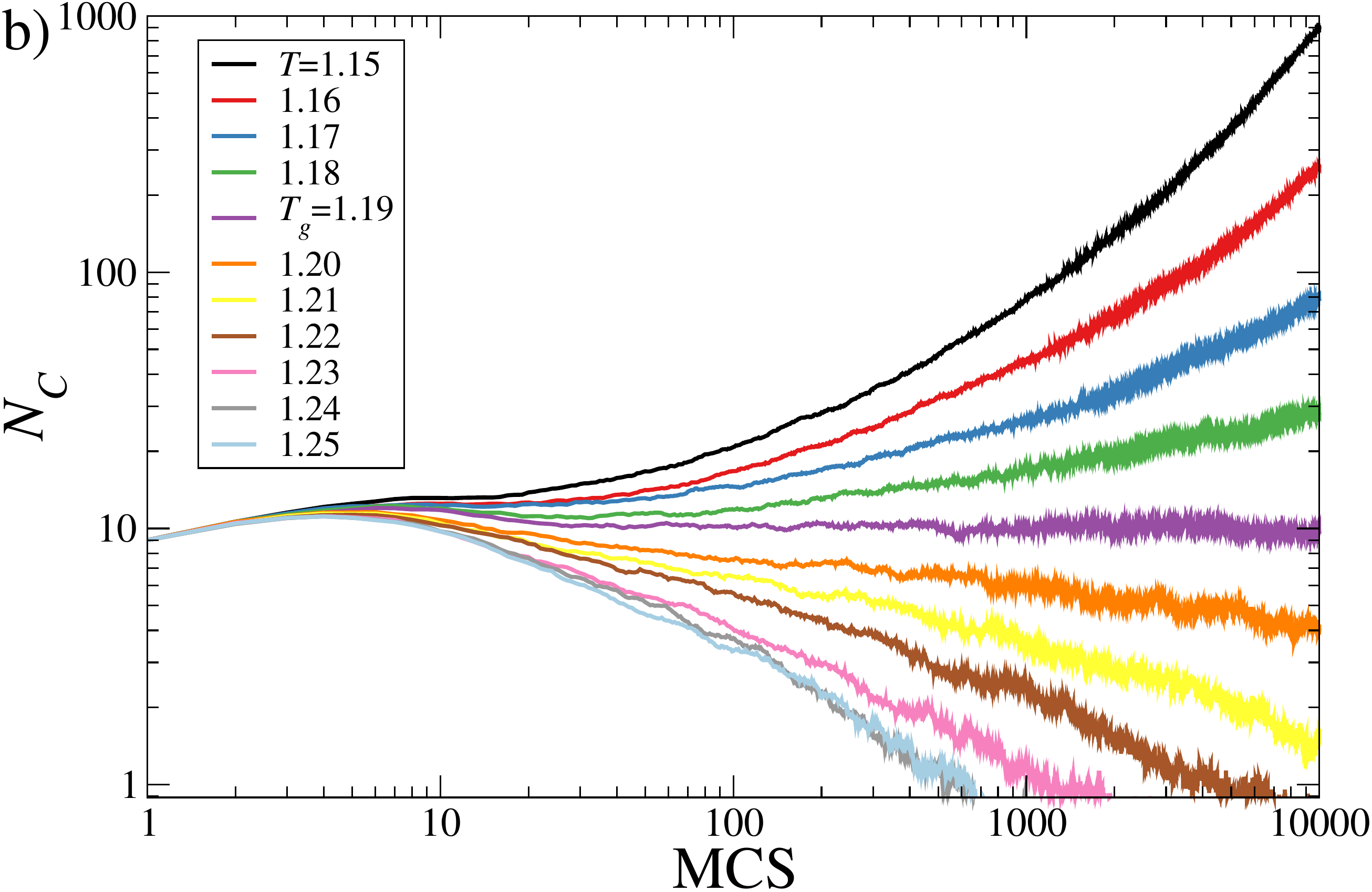}  	
  \caption{Number of cooperators ($N_C$) as a function of time, MCS, for diverse values of temptation, $T$, in a spread dynamic for $1000$ samples in the clean (a) and perturbed model (b). We use a cluster of $3\times3$ cooperators as the initial seed, from where the spread will begin. Here $\Delta=0.3$ for the perturbed model. The clean model exhibits a power-law behaviour in the critical point as $N(t) \propto t^\eta ; \eta=0.243(6)$. The inset shows the average probability of survival for a given spread simulation as a function of time. The critical point for the perturbed model is $T_g<1.19(1)$.}	
  \label{timespread}	
\end{figure*}

Now, we study the dynamics through spreading simulations, that allow us to obtain the critical point in a more precise manner. To do so, we set the initial conditions as a sea of defectors and only a single cooperation seed, made by a $3\times3$ cluster of cooperators. The system size is taken large enough so that activity never reaches the boundary before the end of the simulation. In this setup we are interested only in the initial evolution of the cooperation cluster, and not in the final stable state of the system. We run such dynamics near the critical point, $T_c$, where cooperation is extinct.
For the clean system, at the critical point, we observe a power-law behavior of the mean number of agents, $N(t) \propto t^{-\eta}$. From the data in Figure ~(\ref{timespread}a) we obtain $\eta=0.243(6)$, close to the value $\eta=0.2295(10)$ exhibited by models falling in DP class \cite{Marrobook}. 
We also present the behavior of the average probability of survival for a given spread simulation as a function of time in the inset of Figure \ref{timespread}a). We expect that $P(t)\propto t^{-\delta}$, where we find $\delta=0.42(4)$, in agreement with  the value $\delta=0.4505(10)$ for the DP universality class.

In the disordered system, the critical value $T_g$ is defined as the smallest value supporting asymptotic growth \cite{adr-dic96}. This criterion avoids misinterpretations associated with the effects due to the Griffiths phase, in which power laws in $\rho(t)$ are observed for a range of values of the control parameter \cite{vojta09} in the decay dynamics such as Figure ~\ref{decaydynamics}b). Figure \ref{timespread}b) presents the obtained results for the spreading dynamics. For the perturbed model, the critical point is located near $T_g \simeq 1.19(1)$ when $\Delta=0.3$. We stress that the critical point will change depending on the perturbation strength.

Note that the payoff perturbation is also responsible for lowering the cooperation value when $T<1$. Conversely, this can be seen as increasing the defector's fraction near their extinction point. Based on this observation, we also analyzed how perturbations can induce a symmetric Griffiths phase in the defectors. 
In the perturbed model, for the region around $0.87 < T < 1.18 ~(\Delta= 0.3)$ both cooperators and defectors are in a sub-critical regime, where both quickly reach the equilibrium state and do not fluctuate significantly around the average values.
For $T>1.18$, a clear Griffiths phase appears for the cooperators, where it decays as a power-law with generic exponent, as shown in Figure ~\ref{decaydynamics}b). 
However, we also see that for $T<0.87$ a similar behavior appears for the fraction of defectors when the model is perturbed, that is, the decay in defection is a generic power-law behavior.

\begin{figure*}	
  \includegraphics[width=8cm]{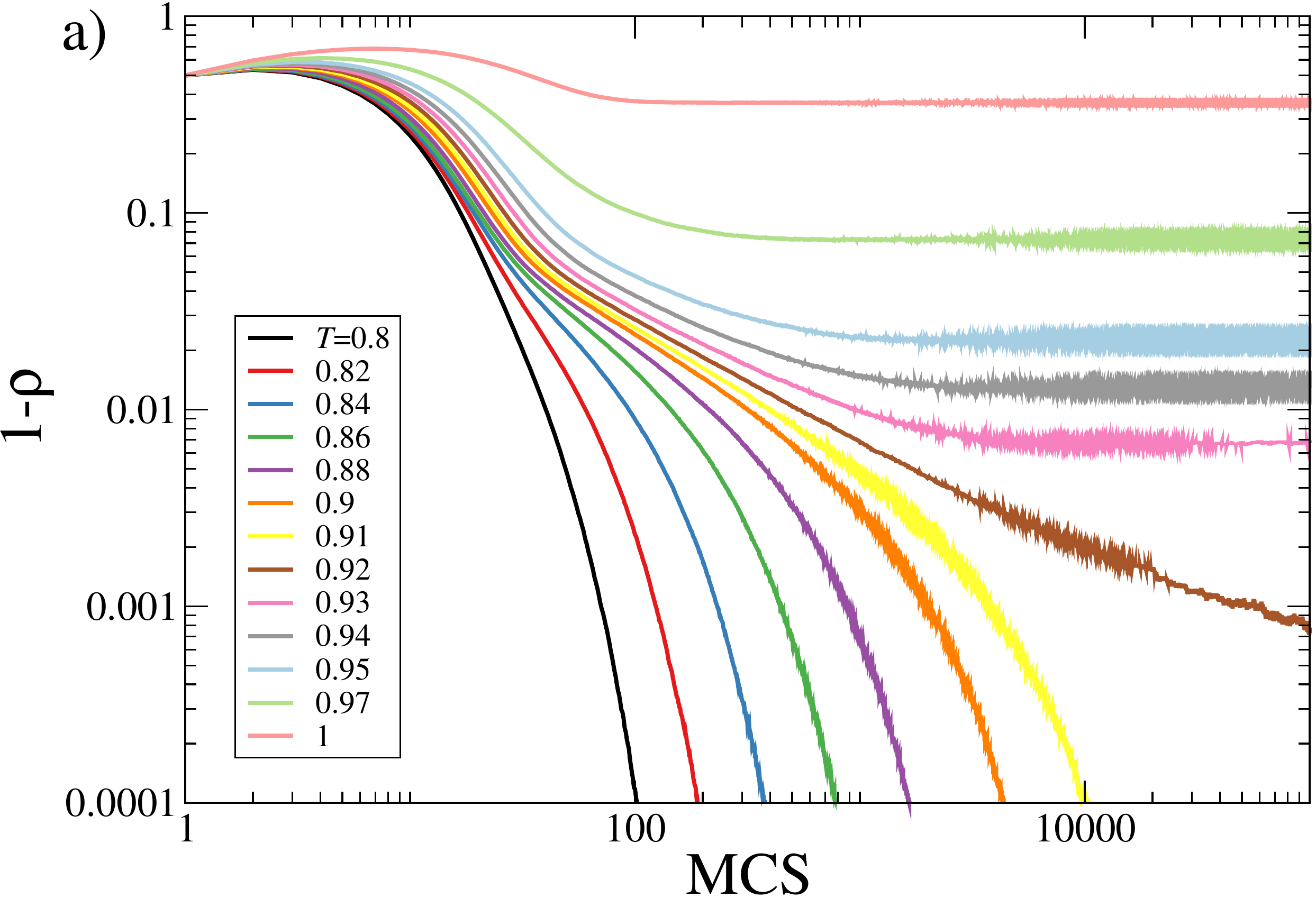}	
  \includegraphics[width=8cm]{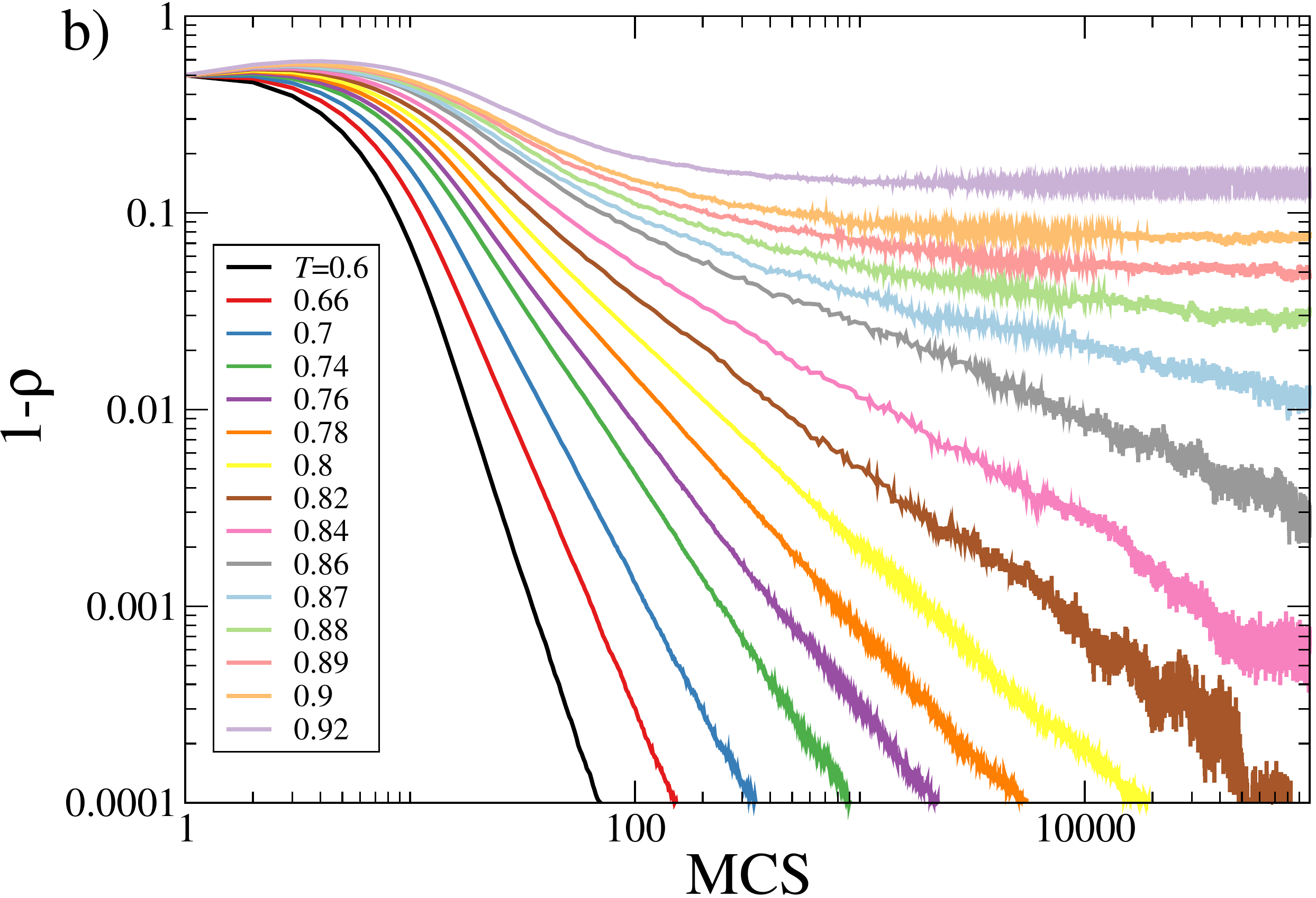}	
    \caption{Time evolution of defectors,$1-\rho$, for the clean (a) and perturbed (b) models near the transition points in the usual population dynamics, i.e. homogeneous starting condition. Here we use $\Delta=0.3$. When looking at the defectors, we can also observe a Griffiths phase in the specific region of $T$ for the perturbed model.}	
  \label{defecttime}	
\end{figure*}

This phenomenon can be seen in Figure ~\ref{defecttime}, which presents the fraction of defectors as a function of time in the region of defectors extinction for the clean, Figure (a), and perturbed model, Figure (b). We set $\Delta=0.3$ for the perturbed model, but general results hold for other perturbation values. We use the typical population dynamics, with homogeneous starting conditions (half the players are cooperators and half are defectors).
The clean model shows the typical stable behavior (for defectors) when $T>0.92$ and an exponential decay otherwise (with an expected power-law decay at the critical point).
On the other hand, for the perturbed model, if $T<0.88$ we can see the power-law decay with a varying exponent but this time for the fraction of defectors. The perturbation is responsible for inducing Griffiths phases in both populations, cooperators, and defectors, for different ranges of the temptation parameter, $T$.
We note that the decay dynamics, starting with $99\%$ of the lattice populated by the strategy that will disappear in the long run, is better for evidencing the Griffiths phase. Nevertheless, even in the usual homogeneous setting, shown in Figure \ref{defecttime}, we can see the power-law decay of the Griffiths phase.

\begin{figure}	
  \includegraphics[width=8cm]{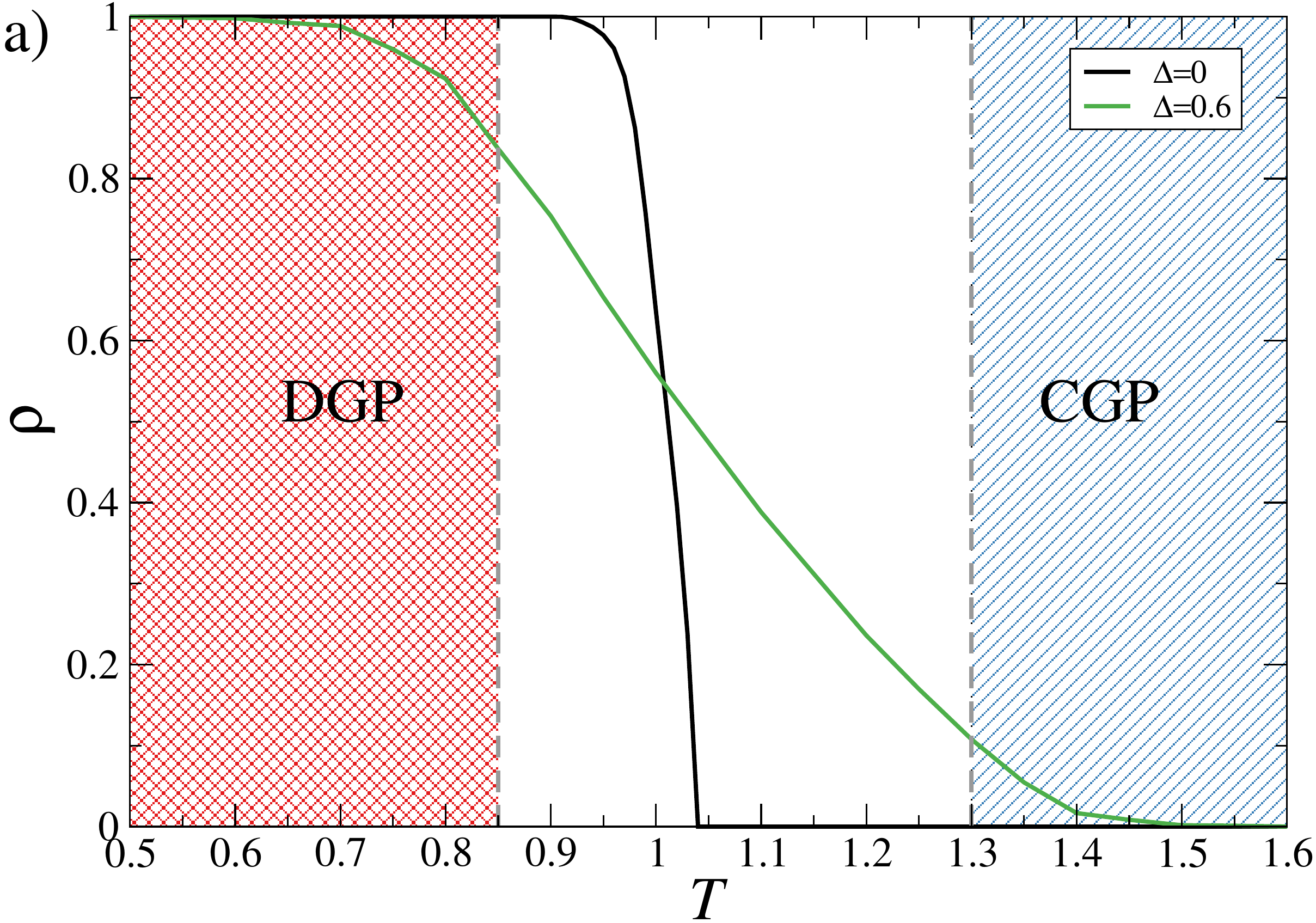}	
  \includegraphics[width=4.2cm]{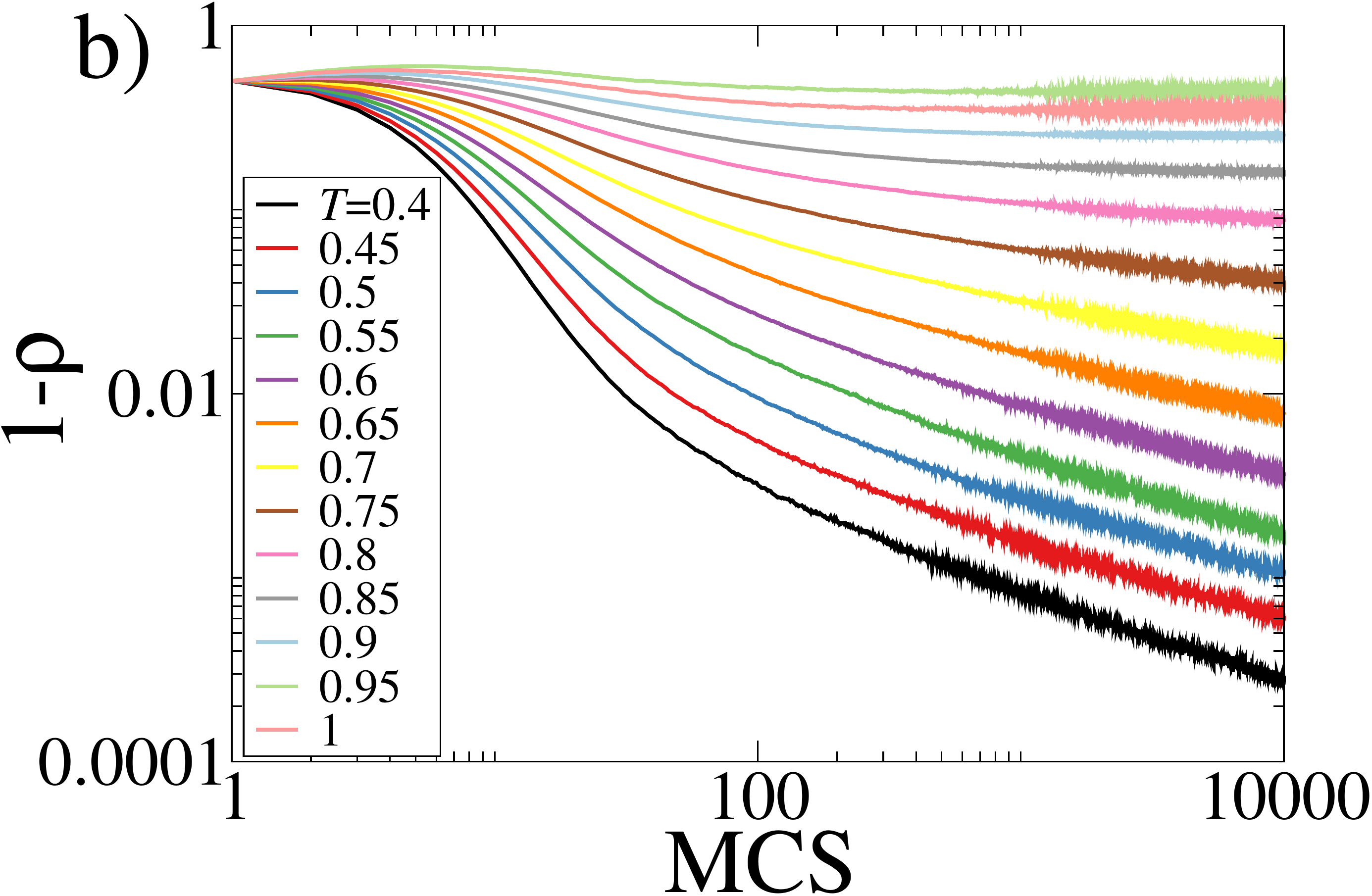}	
  \includegraphics[width=4.2cm]{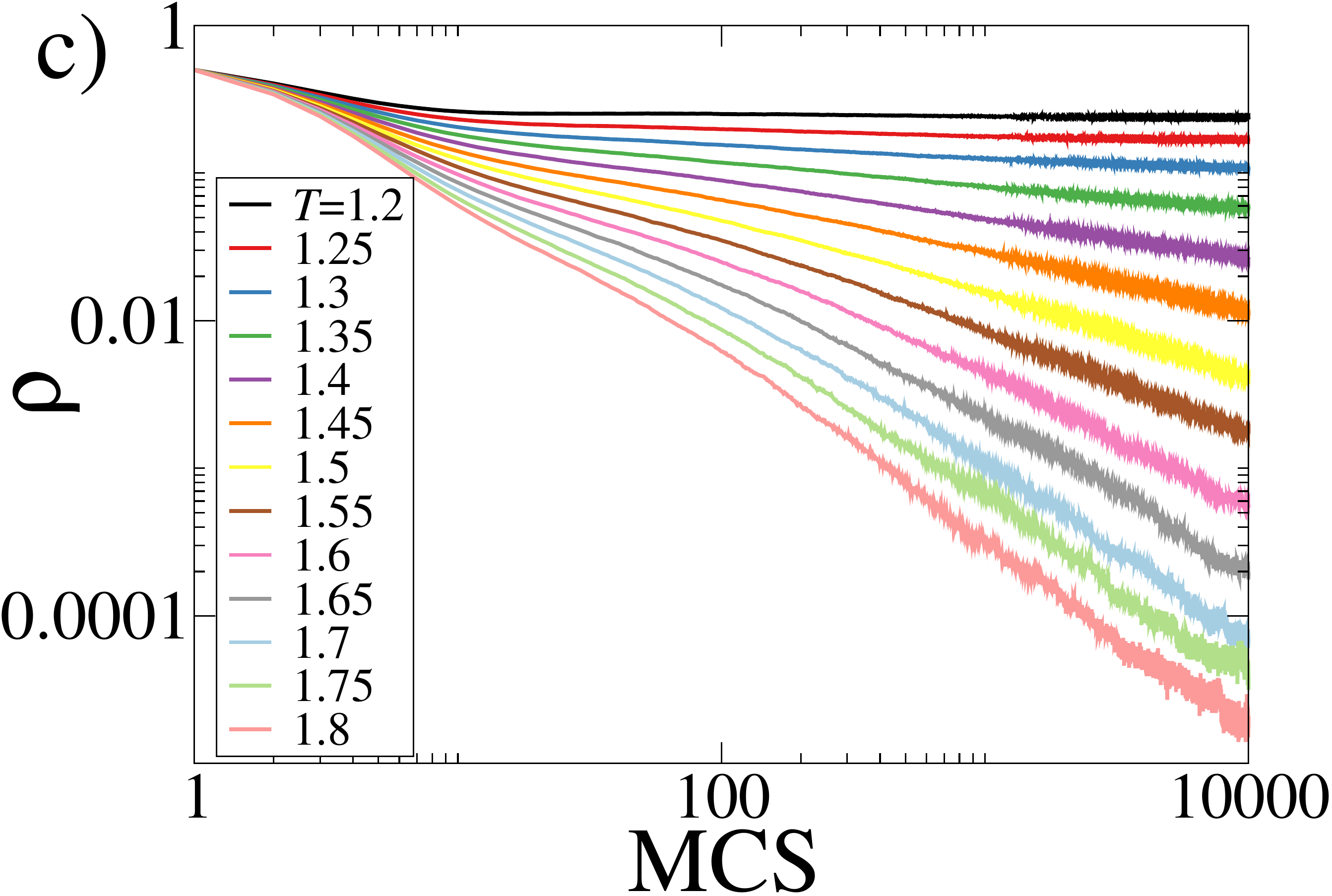}	
  \caption{a)Fraction of cooperators, $\rho$, as a function of $T$ for the strongly perturbed model ($\Delta=0.6$) in comparison with the clean model. For $T<0.85$ there is a Griffiths phase for the defectors (DGP) and another for the cooperators (CGP) when $T>1.3$. The perturbation strength makes the Griffiths phase become broader in the parameter $T$ and more evident in the time evolution of the strategies. Sub-figures b) and c) presents the temporal evolution of the defectors,$1-\rho$, and cooperators respectively. After a given $T$ value, both populations start to decay as a power-law with generic exponent.}	
  \label{grifthsd06}	
\end{figure}

Figure ~\ref{grifthsd06}a) summarizes the results from both observed Griffiths phases. It shows the cooperation fraction, $\rho$, as a function of $T$ for the clean and a strongly perturbed model (we use $\Delta=0.6$ to make more evident the effects of the Griffiths phase). 
We can see that the system has an extended stable phase centered around $T=1$, where both cooperation and defection coexist and quickly reach a stable equilibrium in the dynamics. Nevertheless, near the regions where cooperation, or defection, is extinct we can observe two different Griffiths phases on each end, for each strategy (depicted as cooperator's Griffiths phase, CGP, and defector's Griffiths phase, DGP). 
As we increase the perturbation intensity, such regions get broader. Our results have shown that for weak perturbations ($\Delta=0.1$) there is faint evidence of a Griffiths phase, but the simulation times involved make it prohibitive to verify if this is the case with proper precision. 
For strong perturbations, $\Delta=0.6$, the Griffiths phase is clearly seen for both cooperators and defectors. We present the population temporal evolution in Figure \ref{grifthsd06}b) for the defectors and Figure \ref{grifthsd06}c) for the cooperators. 
In general, the introduction of disorder changes the phase transition of the clean model from a continuous, although steep, decline of cooperation (as a function of $T$) into an almost linear decline for $0.85 < T < 1.3$. The two extreme points for cooperation ($T>1.3$) and defection ($T<0.85$) are the regions where both population dynamics leave the stable regime and start a slow temporal decay as a power law. Note that for infinite times, the graph should present a sharp extinction at these points. Nevertheless, as we have a power-law decay corresponding to the aforementioned Griffiths phases, realistic simulation times make both extreme points smoothly decay into the extinction of cooperation and defection. 

\begin{figure}
  \includegraphics[width=8cm]{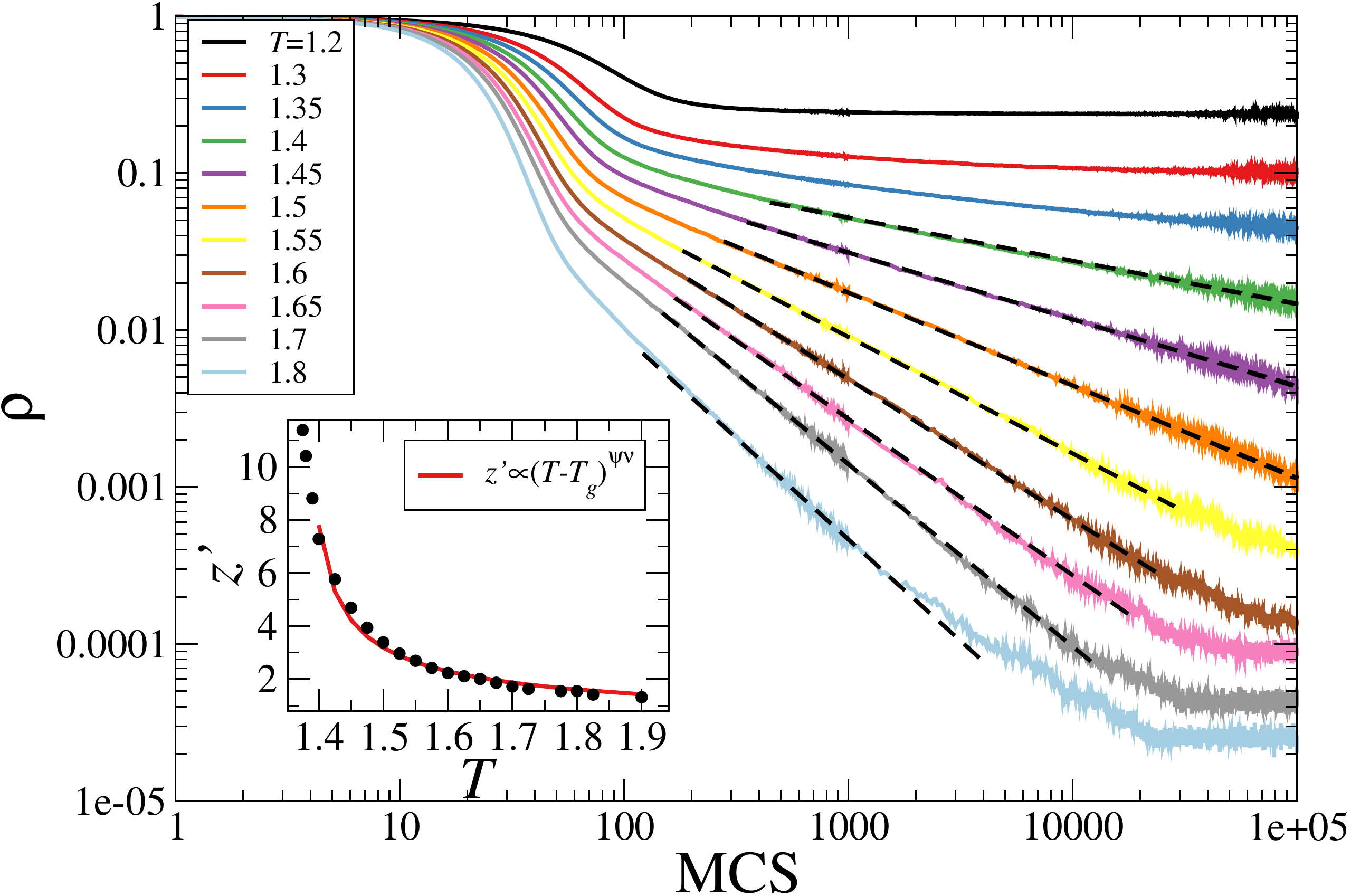}
  \caption{Decay dynamic of the quenched model, using different $T$ values for the strong perturbation dynamics ($\Delta=0.6$). After the critical point ($T_g\simeq1.35$) the decay behaves as a power law with diverse exponents. We can obtain the exponents by fitting the power-law regime as $\rho \propto t^{-2/z'}$. We expect that $z' \propto (T-T_g)^{\psi\nu}$, where $\psi \nu \approx 0.60$. Results from the simulation gives us $\psi \nu = 0.56(5)$. }
  \label{universalc}
\end{figure}

Indeed, in the Griffiths phase scenario, the lifetime of the process follows a power-law decay as $\rho(t)\propto t^{-2/z'}$, with $z'$ being the non-universal dynamical exponent in the Griffiths region \cite{vojta09}, which will depend on perturbation strength $\Delta$ and control parameter $T$. Figure \ref{universalc} presents this analysis for a model with strong perturbation ($\Delta=0.6$) for different values of $T$.  We obtain $z'(T)$ by fitting the power-law decay of the population dynamics for diverse $T$ values.
When approaching the phase transition, $z'$ diverges as $z'\propto |T-T_g|^{-\psi \nu}$, where $\psi$ and $\nu$ are the exponents of the new critical point $T_g$. From the data in Fig. ~\ref{universalc}, we find $\psi \nu=0.56(5)$, consistent with the expected value $\psi \nu \approx 0.60$ of the random transverse Ising model universality class \cite{vojta09}.


Given that the Griffiths phase is mainly characterized by a slow, power-law decay of the order parameter near the phase transitions, a useful measure is the variance of said order parameter, $\sigma^2$. More specifically, let $\sigma^2=<\rho^2>-<\rho>^2$, where $<\rho>$ denotes the cooperation, averaged over the last $N=1000$ Monte-Carlo steps and $300$ different simulations.
During the sub-critical phase, where the population fluctuates around a defined and well-behaved average value, $\sigma^2$ will be small, representing the variance over the average value. Trivially, $\sigma^2$ will tend to zero in the super-critical regime of the clean model, where cooperation will go to zero exponentially.
In the clean model, we expect $\sigma^2$ to have a localized and sharp spike only during the phase transition.
Nevertheless, during a Griffiths phase, $\rho$ will tend to zero in a very slow manner. This effect do not happen only in the exact transition point, instead, we expect $\rho$ to decay as a power law for any $T>T_g$ (e.g. $T_g=1.35$ for $\Delta=0.6$). This results in a fluctuation of $\rho$ around its average value that is larger for a wide range of $T$, when compared with the clean model.

\begin{figure}
  \includegraphics[width=8cm]{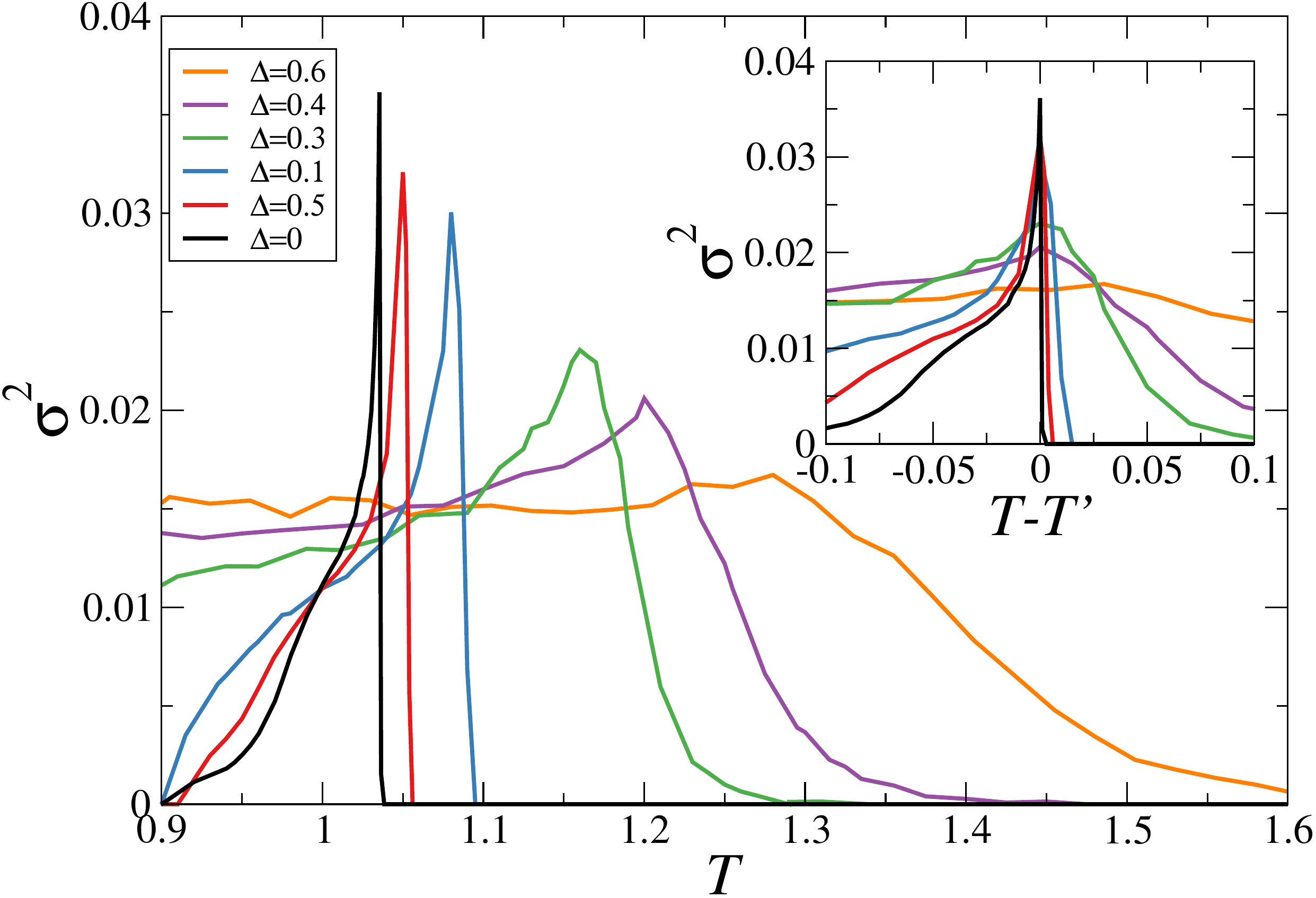}
  \caption{Variance, $\sigma^2$, as a function of $T$ for the clean and perturbed model with different perturbation strengths, $\Delta$. The inset presents the x-axis displaced by $T'$ (the peak position) to make all peaks be centered around $0$. The clean model presents a sharp and localized peak in the variance at the critical point, whereas the perturbed model has a spread in $\sigma^2$ for a wide range of $T$ values. This happens because of the power-law decay observed for a range of the control parameter $T$ during a Griffiths phase. The effects increase continuously with the perturbation strength.}
  \label{sigma}
\end{figure}

In Figure ~\ref{sigma} we present this analysis, showing $\sigma^2$ as a function of $T$ for different perturbation strengths. The inset shows the same information with the x-axis displaced by $T'$, where $T'$ is the position of the peak in $\sigma^2$ for each perturbation. This is done so as to facilitate the comparison of the relative widths of different peaks.
We see that the variance spike in the clean model is strong and very well localized around the transition point (small variations occur due to the finite size and time of the simulations, as expected). On the other hand, the variance for the perturbed model is distributed in a range of $T$ values, where the Griffiths phase is present.
This is another way of observing a possible Griffiths phase in a given model and let us further observe that the characteristics associated with this phase grows continuously with the perturbation strength.
%

\begin{figure}
  \includegraphics[width=4.2cm]{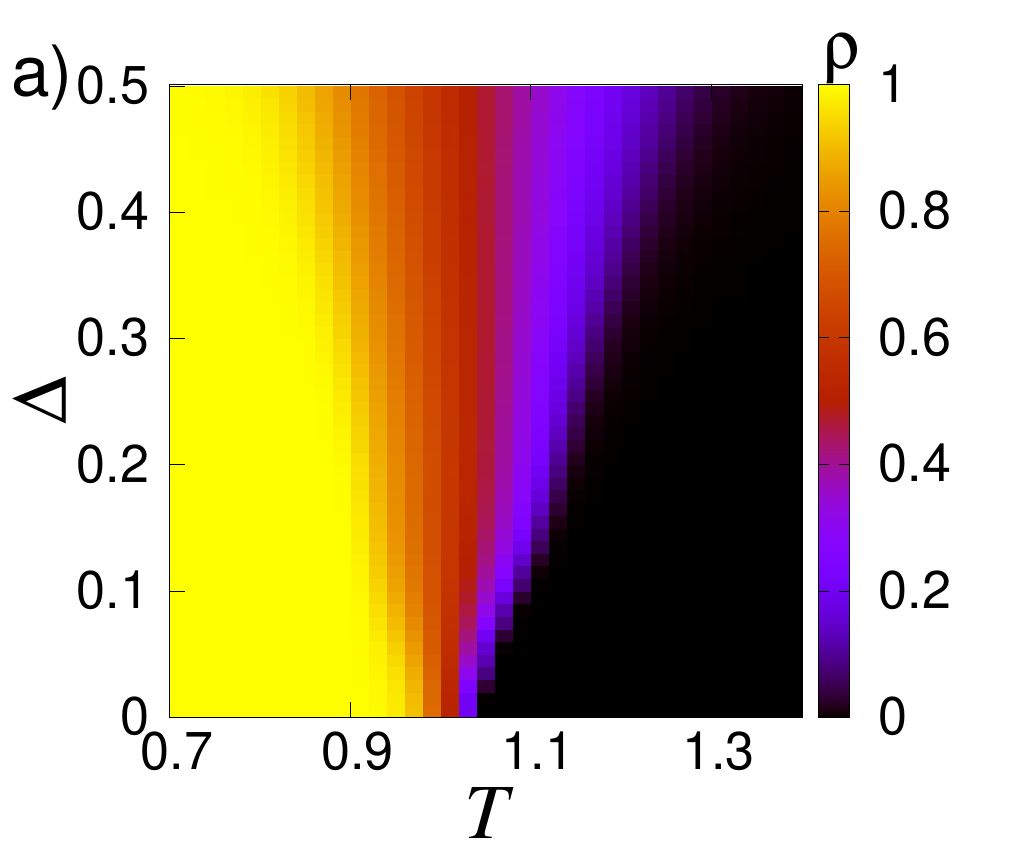}
  \includegraphics[width=4.2cm]{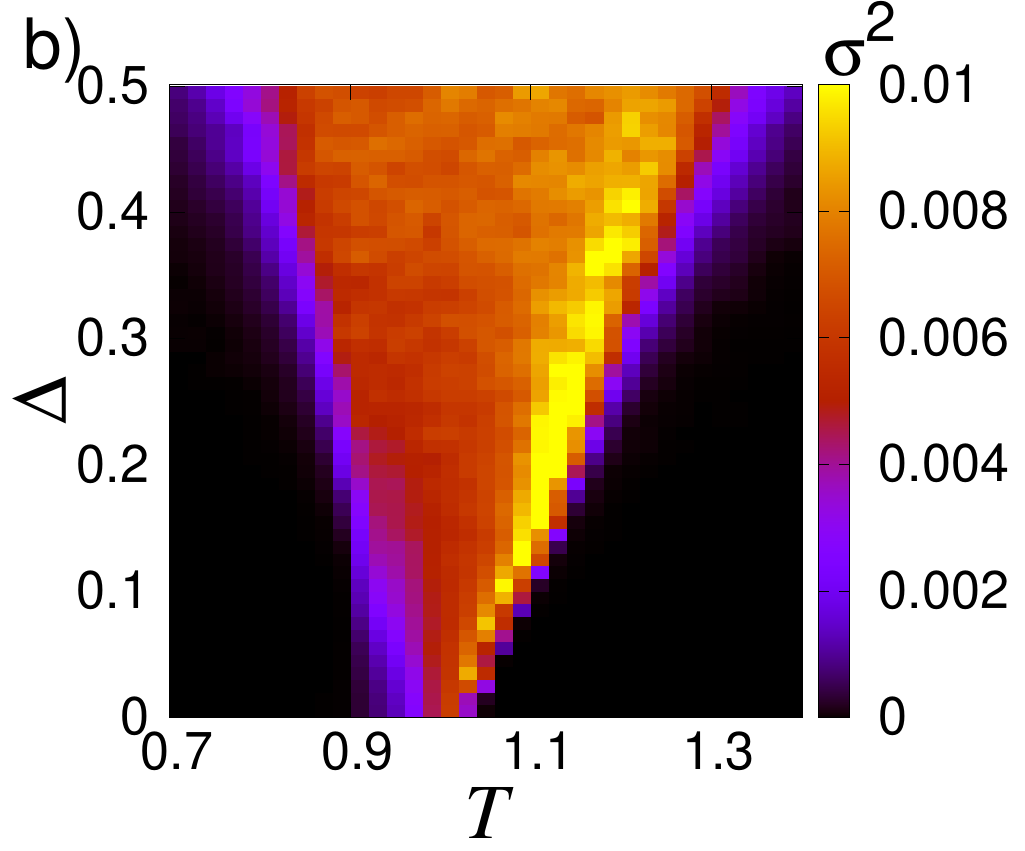}
  \caption{$T \times \Delta $ parameter diagram for a) the final cooperation fraction, $\rho$, and b) the variance, $\sigma^2$.
The increase in $\Delta$ expands the region where cooperation (defection) survives for high (low) $T$.
When the Griffiths phase is present, the variance is distributed along a range of $T$ values, making it an initial indicator of where the Griffiths phase occurs.}
  \label{phase}
\end{figure}

Finally, in figure \ref{phase} we present the final cooperation level in the parameter diagram $T\times \Delta$. As can be seen, the perturbation has the effect of inducing cooperation for $T>1$, while at the same time promoting defection for $T<1$. This is an almost linear effect with the perturbation strength $\Delta$. In summary, the perturbation tends to make the sharp phase transition more smoothly and different strategies able to co-exist. 
We also present in Figure \ref{phase}b) the value of the variance, $\sigma^2$, measured for the last $10^4$ MCS's in a long run of $10^5$ MCS's. This is done so as a way to detect the very slow decay, characteristic of the Griffiths Phase. As expected, we see that the higher $\sigma$ region lies inside the region where perturbation alters the final cooperation fraction and the variance is distributed in a range of $T$, instead of having a single peak in the exact transition point.

\begin{figure}
  \includegraphics[width=4.2cm]{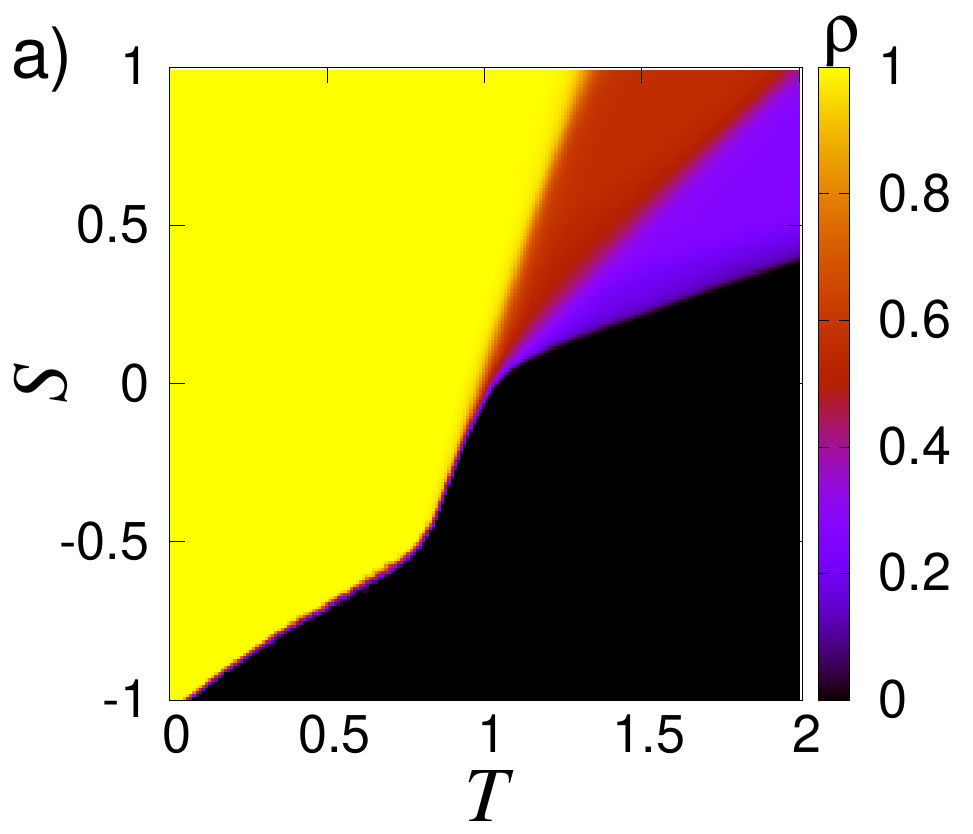}
  \includegraphics[width=4.2cm]{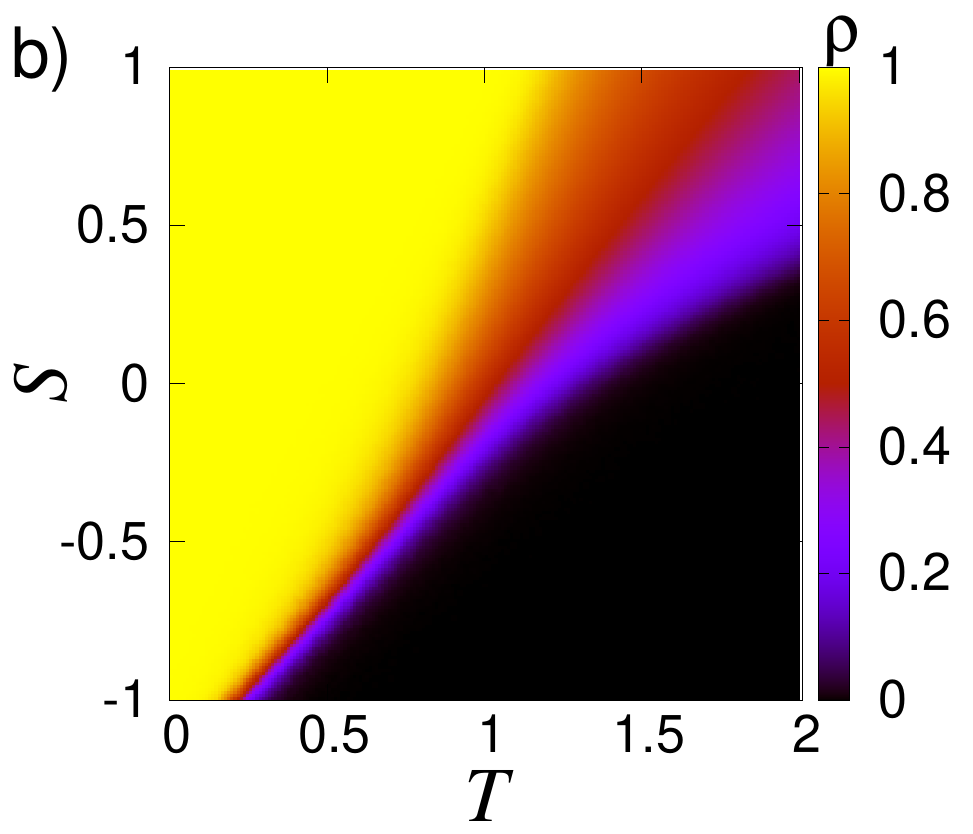}
  \caption{$T \times S$ parameter diagram for the final cooperation level, $\rho$, a) clean model, b) perturbed model with $\Delta=0.6$. The perturbation alters the dynamics mainly near the phase transition regions of the parameter space, making the transition more smooth and continuous.}
  \label{phaserho}
\end{figure}

Finally, we present in Figure \ref{phaserho} the whole $T \times S$ parameter diagram for the final cooperation level, $\rho$, in the clean (a) and perturbed model (b). We fix the perturbation at $\Delta=0.6$ and let the system run for $10^4$ MCS. 
Note that the perturbation sustains the coexistence of strategies near the phase transition for the whole parameter space. In the clean model we have a very sharp transition from full cooperation to full defection in the Stag-Hunt ($S<0,T<1$) and the Prisoners Dilemma game ($S<1,T>1$). On the other hand, the perturbed model have a more smooth transition between those regions, due to the slow power-law decay that allows both strategies to coexist for longer periods of time.

\section{Conclusion}\label{conclusion}

In this work, we have investigated the effects of quenched disorder in the phase transitions of dilemma games through the lens of evolutionary game theory (EGT).
Here, the quenched disorder is included as a small and random perturbation on the payoff matrix of two-player games, representing the typical fluctuations in the perception of risk and reward of a given situation between different individuals.
A common hypothesis in EGT is that all players share the same payoff matrix, and therefore can objectively compare their gains and losses with other players. Nevertheless, in real situations, the perception of a given scenario is much more subjective and can vary due to extrinsic factors. We model such effects as small and random fluctuations, with zero average, around the payoff value.
This gives rise to a quenched (a.k.a. frozen) disorder in the payoff structure of the players. Quenched disorder is a phenomenon widely studied in condensed matter, especially in magnetic medium, since it can give rise to many non-intuitive effects and may drastically alter the properties of a system during phase transitions. Our main goal here was to study how such disorder can affect EGT phase transitions.

Our results reveal that the disorder is able to sustain cooperation in high temptation regions of the prisoner's dilemma game. At the same time, disorder boosts defection for low values of temptation, indicating that such fluctuations tend to allow the coexistence of different strategies for regions where polarization occurs. We also see that the parameter region where both strategies coexist is greatly enhanced when the disorder is present, with its range increasing with the perturbation strength.
However, a more interesting phenomenon is observed when we look at the temporal evolution of the population when the disorder is present. 
The unperturbed (clean) model has a very distinct behavior before and after the classical phase transition point in the parameter $T$ (temptation to defect). Before $T_c$, cooperators behave in a supercritical state, quickly reaching stability, presenting only small fluctuations around a fixed average value. For $T>T_c$ they decay exponentially, quickly reaching extinction. Only when $T=T_c$ we observe a power-law decay (with a unique power-law exponent matching the directed percolation universality class), where cooperators tend to zero in a very slow manner.
On the other hand, we observed that the perturbed model presents a power-law decay for a wide range of $T$ values (instead of only when $T=T_c$), with varying exponents that depend on $T$ and $\Delta$. Such behavior is in contrast with the clean model, where there is only a stable or exponential decay regime. 
This phenomenon is also present for small values of $T$ when we observe how defectors get extinct in the perturbed model.

Such kind of behavior is known as a Griffiths phase, an extended critical-like region that appears in distinct systems, such as disordered magnetic media \cite{magnetic},  epidemic models \cite{epidemics, DeOliveira2008, Odor2015, Cota2016}, and brain networks \cite{brain}. In such an exotic phase, quench disorder can create spatial regions that are locally supercritical even when the parameters of the system are in the sub-critical phase. Such local super-critical stable states prevent the exponential decay near the classical phase transition, creating the power-law decay for a range of the control parameter. In other words, the payoff diversity is able to extend the lifetime of cooperators in a system where cooperation would be extinct otherwise.
Our work shows that Griffiths phases can appear in Evolutionary Game Theory models and it can promote cooperation in regions of high temptation to defect. Even more, we show that this exotic phase may be a common phenomenon in game theory when perception diversity (random payoff perturbations) is present. Given how, in real life, different agents may have different perceptions of the same situation, Griffiths phases may be frequent in social interactions. 
We expect that this work opens a new avenue to study rare phases in evolutionary game theory, especially regarding diverse phenomena that emerge from disordered systems.

\begin{acknowledgments}
This research was supported by the Brazilian Research Agency CNPq (proc. 428653/2018-9) and FAPEMIG.
\end{acknowledgments}

%

\end{document}